\documentclass[12pt]{iopart}
\usepackage{iopams}

\usepackage{epsfig}
\newcommand{\ak}{\ensuremath{a^\dagger}}

\newcommand{\defn}{\stackrel{\rm def}{=}}

\newcommand{\R}{{\mathbb R}}

\newcommand{\im}{{\rm Im\,}}

\begin{document}
\title[Normal forms, inner products, and Maslov indices, of general
multimode squeezings]{Normal forms, inner products, and Maslov indices of general
multimode squeezings}
\author{A M Chebotarev$^1$, %\footnote{The first %author is supported by HSE basic research %programm},
T V Tlyachev$^1$}
%\address{$^1$ Applied Mathematics Department,  %MIEM at National Research University %HSE,109028,Trekhsvjatitelskii per. 3,  Moscow, %Russian Federation}
\address{$^1$ Faculty of Physics, M. V. Lomonosov Moscow State University,
119991, Vorob'evy gory 1, Moscow, Russian Federation}
\ead{chebotarev@phys.msu.ru, tlyachev@physics.msu.ru}
%sadovnikovphysics.msu.ru,
\begin{abstract}
In this paper we present a pure algebraic construction of  the normal factorization of multimode squeezed states and
calculate their inner products. This procedure allows one to orthonormalize bases generated by squeezed states.
We  calculate  several correct representations of  the normalizing constant for the normal factorization, discuss an analogue of the Maslov index for squeezed states, and show that the Jordan decomposition is a useful mathematical tool for
problems with degenerate Hamiltonians. As an application of this theory we consider a non-trivial class of squeezing problems which are solvable in any dimension.

\end{abstract}
\pacs{03.65.Fd, 03.65.-w} %\submitto{Phys. Scripta} \maketitle

\tableofcontents
\pagestyle{empty}
\section{Introduction}
In this paper we  derive a correct expression for the normal ordering of the unitary group
$U_t=e^{i\widehat Ht}$ generated by  the Hamiltonian
\begin{equation}
\label{ham}
\fl
\widehat H=\frac{i}{2}\biggl((\ak,A\ak)-(a,\overline A a)\biggr)+(\ak,B a)+i(\ak,h)-i(a,\overline h)=\widehat H_2+\widehat H_1,
\end{equation}
where $\ak=\{a_{i}^{\dagger}\}_1^n,\;a= \{a_{i}\}_1^n$  are the
multimode creation and annihilation operators with canonical commutation relation (CCR)
$[a_i,\ak_j]=\delta_{ij}$, $A=A^T=\{A_{ij}\}$ is a complex  symmetric
$n\times n$ matrix, $B=B^*=\{B_{ij}\}$ is a Hermitian matrix of the
same size, and $ h\in {\mathbb C}^n$. We use the  standard notation:
the star $B^*$ denotes  the Hermite conjugation of $B$, the bar
$\overline A$ stands for the complex conjugation, and $A^T$ means the
transposed matrix $A$. By $(\,\cdot\,,\,\cdot\,)$ we denote the bilinear
inner product in ${\mathbb R}^n$ and the corresponding bilinear form in
${\mathbb C}^n$; the sesquilinear inner product in the Hilbert state
space ${\cal H}=\otimes_1^n \ell_2$  will be denoted by
$\langle\,\cdot\,,\,\cdot\,\rangle$.

In section 2,  the normal decomposition of generalized
squeezings $U_t=e^{i\widehat Ht}$ is constructed for Hamiltonians \eref{ham} with $A\neq 0$, $B\neq 0$. To this end,  a system of
algebraic equations is derived for $R_t,\,\rho_t,\,C_t\in {\mathbb C}^{n\times
n}$, $g_t,\,f_t\in {\mathbb C}^n$,  and $s_t\in {\mathbb C}$ such that
\begin{equation}
\label{Ut}
\fl
U_t=e^{i\widehat H t}= e^{s_t}e^{-\frac{1}{2} (\ak,R_t\ak)-(g_t,\ak)}\,e^{ (\ak,C_t a)}\,
e^{\frac{1}{2} (a,\overline\rho_t a)+(\overline f_t,a)},\quad U_0=I.
\end{equation}
The solutions are represented in terms of $(n \times n)$-matrices $\Phi_t$ and $\Psi_t$ of canonical transformations preserving
canonical commutation relations  \cite{Be66}. 
Decomposition \eref{Ut} allows one to calculate the normal symbol of squeezing
and the inner products of squeezed states. The last procedure is necessary for
constructing a basis generated by  squeezed states.

For single mode quantum systems, the normal ordered factorization  of
the unitary exponent $U_t=e^{i\widehat H t}$ follows form a formula
proved by D.A.Kirznic in \cite{Ki63}. Applications of this formula to quantum statistics
 are considered in monograph of N.Bogoliubov and D.Shirkov  \cite{Bo82}. The multimode
versions of \eref{Ut}  for $B=0$ was derived by H.-Y.Fan \cite{Fa03}.
For the theory and recent investigations related  to  multimode squeezed
states see the monograph of C.Gardiner and P.Zoller \cite{GaZo04}
and the papers of V.Dodonov  \cite{Do02}, G.Agarwal \cite{SuAg09}, N.Schuch
et al.~\cite{ScCiWo06}.    In \cite{ChRaTl11} we describe the  normal factorization \eref{Ut} of
squeezed states in terms of canonical variables $\Phi_t$ and
$\Psi_t$ introduced  by F.Berezin in \cite{Be66}. We reconsider his proof and suggest new expressions for $s_t$
which preserve the norm of the corresponding squeezed states.

Note that the assumption $B=0$ is typical for the standard definition of
a squeezed state.
The factorization of squeezings  \eref{Ut} with  general matrix $B\neq 0$
was described in \cite{ChRaTl12}. If  $[C_t,\dot C_t]\neq 0$,  difficulties  arise
when one   tries  to  derive an evolution equation for  $C_t$
in decomposition \eref{Ut} (see \cite{Fe51}
and \cite{KaMa91}, pp. 274--275, Eq.~(1.10)). The advantage of
canonical variables $\Phi_t$, $\Psi_t$
 is that they allows one to derive and to solve just {\it algebraic} equations for matrices
 $R_t,\;C_t,\;\rho_t$ in \eref{Ut}, but not a nonlinear  ODE, which can not be
 written for $C_t$  as a local ODE, when $[C_t,\dot C_t]\neq 0$ (see \cite{ChRaTl12}).

A short proof of the normal factorization
\eref{Ut} and explicit representations of the matrix valued coefficients for this
decomposition in terms of canonical transformations are considered in section 2.

In section 3, we derive  integral representations
for the scalar function $s_t$  which defines the norm and the phase of the normal
decomposition and discuss the index problem, which is essential for
systems with $B\neq 0$ and implies continuity of $s_t$. The algebraic representations of $s_t$
can be calculated faster than the corresponding
integral expressions.

Algebraic expressions for $s_t$ and  the formula for
the normal symbol of squeezings are discussed in section 4.

In section 5, we recall some useful facts on $L_2(\R^n)$-representations of multimode
squeezings and
establish equations representing the inner product of squeezings and
compositions of squeezed states. In this way, the orthonormalization procedure
for  squeezed states can be reduced to standard problems of linear algebra.

The algebraic expressions for components of the Jordan decomposition
of matrces generating the canonical transformations are derived in section 6 . This procedure is helpful for solving the problems with degenerate  Hamiltonians.

In section 7, we note that in the class of problems with
$A$ and $B$ such that $[B,A\overline A]=0$, the
factorization problem reduces to the eigenvalue problem for the Hermitian matrix
$A\overline A - B^2$.

Numerical tests are considered in section 8.
The basic equalities have been checked either analytically or numerically by using Wolfram Mathematica, and
these  interactive tests are available at \cite{statphys}.

\section{Canonical transformations and normal representation of squeezings}
Hamiltonian   \eref{ham}  defines the $(2n\times 2n)$-block matrix
{\small $G = \left(\begin{array}{cc}
                 -iB&A\\
                 \overline A&i\overline{B}
\end{array}\right)$} and the group of symplectic matrices $e^{Gt}$
 (see \cite{Be66}, \cite{Go13}) such that
\begin{eqnarray}
\label{defG}
\fl
i\biggl[\widehat H,\left(\begin{array}{c}
         a \\
         \ak
\end{array}\right)\biggr]=
G
\left(\begin{array}{c}
              a\\
                \ak
\end{array}\right)
+
\left(\begin{array}{c}
              h\\
                \overline h
\end{array}\right), \quad
S_t\defn e^{Gt}=\left(\begin{array}{cc}
                 \Phi_t&\Psi_t\\
                 \overline \Psi_t&\overline\Phi_t
\end{array}\right),\quad t\in {\mathbb R}.
\end{eqnarray}
The matrices $S_t$ preserve $(2n\times 2n)$-block structure \eref{defG} and possess the following
properties: ${\rm det\,} S_t=1$,
\begin{eqnarray}
\label{inversion}
\fl
S_{-t}=S_{t}^{-1}=\left(\begin{array}{cc}
                 \Phi_{-t}&\Psi_{-t}\\
                   \overline\Psi_{-t}&\overline \Phi_{-t}
\end{array}\right)=
\left(\begin{array}{cc}
                 \Phi^*_{t}&-\Psi^T_{t}\\
                  -\Psi^*_{t}&\Phi^T_{t}
\end{array}\right),\;
S_t^TJS_t=J,\quad   J=\left(\begin{array}{cc}
                 0&I\\
                  -I&0
\end{array}\right).
\end{eqnarray}

Equations \eref{defG} define the evolution 
$\dot S_t
 =S_tG =GS_t$, initial values  $\Phi_0=I$,  $\Psi_0=0$,
%\begin{eqnarray}
%\label{sys2}
%\label{commEq}
%\fl
%\dot S_t=G
% =S_tG =GS_t,
%      \end{eqnarray}
and algebraic representations
for  $a_t=U_t a U_t^*$, $\ak_t=U_t \ak U_t^*$, $h_t$,  and $\overline
h_t$:
\begin{eqnarray}
\label{sys1}
\fl
\left(\begin{array}{c}
                 a_t\\
                 \ak_t
 \end{array} \right)=
S_t\left(\begin{array}{c}
                 a\\
                 \ak
\end{array}\right)+
\left(\begin{array}{c}
                 h_t\\
                 \overline h_t
 \end{array} \right),
 \quad
 \left(\begin{array}{c}
                 h_t\\
                 \overline h_t
 \end{array} \right)
 \stackrel{\rm def}{=}
 \int_0^t S_\tau \left( \begin{array}{c}
                 h\\
                 \overline h
\end{array} \right)d\tau
      =\frac{S_t-I}{G}\left( \begin{array}{c}
                 h\\
                 \overline h
\end{array} \right).
\end{eqnarray}
The matrices
\begin{equation*}
\fl
G^{-1}(\exp{Gt}-I)=I+\frac{1}{2!} G+\frac{1}{3!}G^2+\dots, \quad
G^{-2}(\exp{Gt}-I-Gt)=\frac{1}{2!}I +\frac{1}{3!}G
+\dots
\end{equation*}
remain well defined for degenerate $G$.

The set of canonical commutation
relations and the rules for inversion of time
\begin{eqnarray}
\label{CR}
\fl
\Phi_{t}\Phi^*_{t}-\Psi_{t}\Psi^*_{t}=\Phi^*_{t}\Phi_{t}-\Psi^T_{t}\overline \Psi_{t}=I,
\quad
\Phi_{t}\Psi^T_{t}-\Psi_{t}\Phi^T_{t}=\Phi^*_{t}\Psi_{t}-\Psi^T_{t}\overline \Phi_{t}=0,
\\
\label{invT}
\fl
R_t=-\rho_{-t},\quad \Phi_t=\Phi^*_{-t},\quad,\Psi_t=-\Psi^T_{-t}
\end{eqnarray}
is a corollary of equations \eref{inversion} and the identity
$S_tS_{-t}=S_{-t}S_{t}=I$.
More generally, from $e^{G(t\pm s)}=e^{Gt}e^{\pm Gs}=e^{\pm Gs}e^{Gt}$ and \eref{inversion}, the matrix analogue of  addition-subtraction formulae for sine and cosine follow:
\begin{eqnarray}
\label{compose}
\fl
\nonumber
\Phi_{t+s}=\Phi_t\Phi_s+\Psi_t\overline \Psi_s=\Phi_s\Phi_t+\Psi_s\overline\Psi_t,\quad
\Psi_{t+s}=\Psi_t\Phi_s+\Phi_t\overline \Psi_s=\Phi_s\Psi_t+\Psi_s\overline\Phi_t,
\\
\fl
\Phi_{t-s}=\Phi_t\Phi^*_s-\Psi_t\Psi^*_s=\Phi^*_s\Phi_t-\Psi_s^T\overline\Psi_t,\quad
\Psi_{t-s}=\Psi_t\Phi^T_s-\Phi_t\Psi^T_s=\Phi^*_s\Psi_t-\Psi_s^T\overline\Phi_t.
\end{eqnarray}

Identities \eref{CR} imply the inequality $\Phi_t\Phi_t^*\ge I$, so that
the inverse matrix $|\Phi_t|\ge I$, i.e. $\Phi_t^{-1}$ exists. The second
identity \eref{CR} proves that the matriñes
\begin{equation*}
\fl
R_t=\Phi_t^{-1}\Psi_t=\Psi^T_t(\Phi_t^T)^{-1}=R_t^T,\quad \rho_t=\Psi_t \overline{\Phi}_t^{-1}=
(\Phi_t^*)^{-1}\Psi_t^T
\end{equation*}
are  {\it symmetric} and
 {\it well  defined} for all $t\ge 0$. Moreover, equation \eref{CR} implies
\begin{equation*}
\fl
R_t\overline R_t= R_tR^*_t=I-(\Phi_t^*\Phi_t)^{-1}\le I,  \quad |R_t|^2=R^*_tR_t\le I.
\end{equation*}
Therefore, the operator $e^{\pm\frac12(\ak,R_t \ak)}$ in \eref{Ut} is
densely defined at any time $t\in{\mathbb R} $.

If $A=\overline A=0$, then
{\small
\begin{eqnarray*}
\fl
e^{Gt}=\left(\begin{array}{cc}
                 \Phi_t&0\\
                 0&\overline \Phi_t
\end{array}\right)=\left(\begin{array}{cc}
                 e^{-itB}&0\\
                 0&e^{it\overline B}
\end{array}\right)
\end{eqnarray*}
}
and the unitary group $U_t=e^{it\widehat H}$ can be rewritten as a normally ordered composition
\begin{eqnarray*}
\fl
e^{it\widehat H_2}=e^{s_t} e^{-g_t,\ak} e^{(\ak,C_t a)} e^{(\overline f_t,a)}=e^{s_t} e^{-g_t,\ak} :e^{(\ak,(e^{C_t}-I) a)}: e^{(\overline f_t,a)},
\end{eqnarray*}
where the creation and annihilation operators inside the colon brackets act in the normal order.
Explicit equations for $s_t$, $g_t$, $f_t$, and $C_t$ readily follow from CCR and the pair of equivalent representations of $a_t=U_taU^*_t$ and
 $\ak_t=U_t\ak U^*_t$:
\begin{eqnarray*}
\fl
\Phi_t a+h_t=e^{it\widehat H}ae^{-it\widehat H}= e^{-(g_t,\ak)} e^{(\ak,C_t a)} ae^{-(\ak,C_t a)} e^{(g_t,\ak)}=e^{-C_t}(a+g_t),\\
\fl
\overline\Phi_t \ak+\overline h_t=e^{it\widehat H}\ak e^{-it\widehat H}= e^{(\ak,C_t a)}e^{(\overline f_t,a)} \ak e^{-(\overline f_t,a)}e^{-(\ak,C_t a) }=e^{C^T_t}\ak+\overline f_t.
\end{eqnarray*}
By equating the coefficients at $a$, $\ak$, and at the operator of multiplication by scalar on the left and right hand sides of these equalities, we obtain
\begin{eqnarray}
\label{A0}
\fl
g_t=\Phi_t^{-1}h_t,\; f_t=h_t,\;C_t:\; \Phi_t=e^{-itB}=e^{-C_t},\;
C_t=itB.
\end{eqnarray}
In order to calculate $e^{s_t}=\langle 0|e^{it\widehat H}|0\rangle$, one can use the following equation:
\begin{eqnarray}
\fl
\label{commA0}
\dot s_te^{s_t}=\langle 0|e^{it\widehat H}i\widehat H|0\rangle
=-\langle 0|e^{it\widehat H}(h,\ak)|0\rangle=
-\langle 0|(h,\overline\Phi_t \ak+\overline h_t)e^{it\widehat H}|0\rangle=-e^{s_t}(h,\overline h_t).
\end{eqnarray}
Therefore, $s_t=-\int_0^t(h,\overline h_\tau)d\tau$, and
finally we obtain the normal decomposition for squeezings with $A=0$:
\begin{eqnarray}
\fl
\label{expA0}
e^{it\widehat H}=e^{-\int_0^t(h,\overline h_\tau)d\tau} e^{-(\Phi_t^{-1}h_t,\ak)}:e^{(\ak,(e^{itB}-I)a)}:e^{(\overline h_t,a)}.
\end{eqnarray}

For Hamiltonian \eref{ham}, the general form of the normal decomposition
\begin{eqnarray}
\label{genDec}
\fl
e^{it\widehat H}=e^{s_t} e^{-\frac12(\ak,R_t\ak)-(g_t,\ak)} :e^{(\ak,(e^{C_t}-I) a)}: e^{\frac12(a,\overline \rho_t a)+(\overline f_t,\ak)}.
\end{eqnarray}
is more similar to an
expression used for $B=0$ (see  \cite{ChRaTl11}):
{\small
\begin{eqnarray*}
\fl
e^{Gt}=e^{t\left(\begin{array}{cc}
                 0&A\\
                 \overline A&0
\end{array}\right)}
=\left(\begin{array}{cc}
                 \Phi_t&\Psi_t\\
                 \overline \Psi_t&\overline \Phi_t
\end{array}\right)
,\quad
\Phi_t=\Phi^*_t=\cosh(A\overline A)^{\frac12} t,\quad
\Psi_t=\Psi_t^T=\frac{\sinh(A\overline A)^{\frac12}t}{(A\overline A )^{\frac12}}A.
\end{eqnarray*}
}

The proof of \eref{genDec} in the general case uses the commutation rules
\begin{eqnarray*}
\fl
e^{-\frac12(\ak,R_t\ak)-(g_t,\ak)}  a e^{\frac12(\ak,R_t\ak)+(g_t,\ak)}=a+R_t\ak+g_t,
\\
\fl
e^{\frac12(a,\overline\rho_t a)+(\overline f_t,a)}  \ak e^{-\frac12(a,\overline\rho_t a)-(\overline f_t,a)}=
\ak +\overline \rho_t a+\overline f_t
\end{eqnarray*}
and the equations for
parameters of the normal decomposition which follow from the commutation relations
{\small
\begin{eqnarray}
\fl
\nonumber
\Phi_t a+\Psi_t+h_t=e^{it\widehat H}ae^{-it\widehat H}
\\
\label{canTrans}
\fl
=e^{-\frac12(\ak,R_t\ak)-(g_t,\ak)} e^{(\ak,C_t a)} ae^{-(\ak,C_t a)} e^{\frac12(\ak,R_t\ak)+(g_t,\ak)}=e^{-C_t}(a+R_t\ak+g_t),
\\
\nonumber
\\
\fl
\nonumber
\overline\Phi_t \ak+\overline\Psi_t a+\overline h_t=e^{it\widehat H}\ak e^{-it\widehat H}
\\
\fl
\nonumber
=e^{-\frac12(\ak,R_t\ak)-(g_t,\ak)}e^{(\ak,C_t a)} (\ak +\overline \rho_t a+\overline f_t) e^{-(\ak,C_t a) }e^{\frac12(\ak,R_t\ak)(g_t,\ak)}
\\
\nonumber
\fl
=e^{-\frac12(\ak,R_t\ak)-(g_t,\ak)}(e^{C_t^T}\ak +\overline \rho_t e^{-C_t} a+\overline f_t) e^{\frac12(\ak,R_t\ak)(g_t,\ak)}
\\
\nonumber
\fl
=e^{C_t^T}\ak +\overline\rho_t e^{-C_t}(a+R_t\ak+g_t)+\overline f_t.
\end{eqnarray}
}
\noindent
These relations imply
equations for  parameters $R_t$, $\rho_t$, $C_t$, $g_t$, $h_t$ of the normal decomposition \eref{Ut}:
\begin{eqnarray}
\label{baseEq}
\fl
\Phi_t=e^{-C_t},
\quad \overline\Phi_t=e^{C^T_t}+\overline \rho_t e^{-C_t}R_t,\quad h_t=e^{-C_t}g_t,
\\
\fl
\nonumber
\Psi_t=e^{-C_t}R_t,\quad\overline \Psi_t=\overline \rho_t e^{-C_t},\quad \overline h_t=
\overline \rho_t e^{-C_t}g_t+\overline f_t.
\end{eqnarray}
This system of equations possesses   the following   solution:
\begin{eqnarray}
\fl
\label{solEq}
C_t=-\ln \Phi_t,\quad R_t=\Phi_t^{-1}\Psi_t,\quad
\overline\rho_t=\overline\Psi_t\Phi_t^{-1}, \quad g_t=\Phi_t^{-1}h_t,
\quad f_t=h_t-\rho_t\overline h_t.
\end{eqnarray}
The compatibility of  equations \eref{baseEq} for $\Phi_t$ and $\overline \Phi_t$, $\Psi_t$ and
$\overline \Psi_t$ is a  remarkable fact:
{\small
\begin{eqnarray*}
\fl
e^{C^T_t}+\overline \rho_t e^{-C_t}R_t=(\Phi_t^T)^{-1}+\overline\Psi_t
\Phi_t^{-1}\Psi_t=(\Phi_t^T)^{-1}+(\Phi_t^T)^{-1}\Psi^*_t\Psi_t
\\
\fl
=
(\Phi_t^T)^{-1}+(\Phi_t^T)^{-1}(\Phi^T_t\overline \Phi_t-I)=\overline \Phi_t,
\quad
\overline{\Psi_t}=e^{-\overline{C}_t}\overline{R}_t=\overline{\Psi}_t\Phi_t^{-1}\Phi_t=\overline{\rho}_te^{-C_t}.
\end{eqnarray*}
}
Thus, the following theorem is proved.
\medskip

\begin{thm} {\bf 1}.
{\it
The  vector-valued   and matrix-valued
coefficients of the normal decomposition \eref{Ut} of the squeezing   with Hamiltonian \eref{ham}
\begin{equation}
\label{coeffic}
\fl
R_t=\Phi_t^{-1}\Psi_t, \quad \overline \rho_t=\overline\Psi_t\Phi_t^{-1}, \quad C_t=-\ln \Phi_t, \quad g_t=\Phi_t^{-1}h_t, \quad f_t
 = h_t- \rho_t\overline h_t
 \end{equation}
are well defined in terms of $\Phi_t$ and $\Psi_t$ by \eref{defG}.
The matrices $\Phi_t^{-1}$ and
$G^{-1}(e^{Gt}-1)$ are well defined for any given $A=A^T$, $B=B^*$ at any time  $t\in \mathbb R$.
}
\end{thm}

\medskip

\section{Integral representations of $s_t$ and the index problem}
Let us  calculate $e^{s_t}$ by using the vacuum expectation $ e^{s_t}
=\langle 0|e^{i\widehat H t}|0\rangle$ (see \eref{Ut}) and one of the two obvious equations: $\dot s_t
=ie^{-s_t}\langle 0|e^{i\widehat H t}\widehat H|0\rangle$ or $\dot s_t
=ie^{-s_t}\langle 0|\widehat H e^{i\widehat H t}|0\rangle$.

By
definition of the vacuum state, we have
\begin{eqnarray*}
\fl
\langle 0|e^{i\widehat H t}(a,\overline h)|0\rangle=0,
\quad
\langle 0|e^{i\widehat H t}(\ak, B a)|0\rangle=0,
\quad
\langle 0|e^{i\widehat H t}(a,\overline A a)|0\rangle=0.
\end{eqnarray*}
Definition \eref{solEq} of
$\overline\rho_t$ and canonical transformations \eref{canTrans} justify the relationship
\begin{equation}
\label{useful}
e^{i\widehat H
t}(\ak-\overline\rho_t a)\,e^{-i\widehat Ht}=(\Phi_t^T)^{-1} \ak+\overline f_t.
\end{equation}
As a corollary of
\eref{useful} we find the two basic  vacuum expectations:
\begin{eqnarray*}
%\label{eqComm}
\fl
\langle 0|e^{i\widehat H t}(\ak,h)|0\rangle=\langle 0|e^{i\widehat H t}((\ak-\overline\rho_t a),h)|0\rangle=
 \langle 0|((\Phi_t^T)^{-1} \ak+ \overline f_t,h)e^{i\widehat H t}|0\rangle=e^{s_t}
(\overline f_t,h),
\\
\fl
\langle 0|e^{i\widehat H t}(\ak,A\ak)|0\rangle=
\langle 0|e^{i\widehat H t}\bigl((\ak-\overline\rho_t a),A(\ak-\overline\rho_t a) \bigr)|0\rangle+\langle 0|e^{i\widehat H t}|0\rangle \,{\tr\,}\overline\rho_t A
\\
\fl
\quad =
\langle 0|(\overline f_t,A\overline f_t)e^{i\widehat H t}|0\rangle +e^{s_t}{\tr\,}\overline\rho_t A
=e^{s_t}\bigl((\overline f_t,A\overline f_t)+{\tr\,}\overline\rho_t A \bigr).
\end{eqnarray*}
Therefore,
$\dot s_t=ie^{-s_t}\langle 0|e^{i\widehat H t}\widehat H|0\rangle=-(\overline f_t,h)-\frac12\biggl( (\overline f_t,A\overline f_t) +{\tr\,}\overline\rho_t A\biggr)$ and
this equality  proves Lemma 2.

\medskip

{\bf Lemma 2.}
{\it
For $f_t$ and $\rho_t$ defined by theorem 1, we have
\begin{eqnarray}
\label{eqS}
\fl
 s_t=-\int_0^t \biggl((\overline f_\tau,h)+
\frac12\,(\overline f_\tau,A\overline f_\tau)+\frac12\,\tr\,\overline\rho_\tau A
\biggr)\,d\tau, \quad \overline f_t=\overline h_t-\overline\rho_t h_t.
\end{eqnarray}
}

If $A=0$, then $\rho_t=0$, $f_t=h_t$, and \eref{commA0} coincides with  function \eref{eqS}.

An equivalent representation of $s_t$ follows from  $\dot s_t
=ie^{-s_t}\langle 0|\widehat H e^{i\widehat H t}|0\rangle$, \eref{inversion},
and the equality  $e^{-it\widehat H}
(a+R_t\ak)e^{it\widehat H}=-(\Phi_{-t}^*)^{-1}a+h_{-t}-\rho_{-t}\overline h_{-t}=
\Phi_t^{-1}a+f_{-t}$:
\begin{eqnarray}
\label{eqS2}
\fl
 s_t=\int_0^t \biggl((\widetilde f_{\tau}, \overline h)+
\frac12\,(\widetilde f_{\tau},\overline A\widetilde  f_{\tau})-\frac12\,\tr\,R_\tau \overline A
\biggr)\,d\tau,\quad  \widetilde f_{t}=h_{-t}+R_t\overline h_{-t}.
\end{eqnarray}
Equivalence of \eref{eqS}  and \eref{eqS2}  was also tested numerically for randomly simulated $A$, $B$, and $h$ (see \cite{statphys}).

The expression for  $s_t$  in Berezin's book  (see  \cite{Be66}, (6.24) in p.~143)
differs from \eref{eqS} and \eref{eqS2}.   Taking into account the correspondence
of notations,
\begin{table}[h!]
\caption{}
\vskip1mm
\label{BasicDistr}
\begin{tabular}{|c|c|c|c|c|c|c|}
\hline
Ber &$A$ & $i\overline A$ & $C$ & $f$ & $i\overline f$& $g_t$\\
\hline
Che-Tl & $iA$ & $\overline A$ & $B$& $ih$&  $\overline h$& $h_t$ \\
\hline
\end{tabular}
\end{table}
%\begin{figure}[h!]
%\centering\epsfig{figure=berNot,width=0.4\linewidth}
%\end{figure}
his expression of the normalizing factor is equal to
\begin{eqnarray}
\label{eqBe}
\fl
e^{s_t^{({\rm Be})}}=
\frac{e^{-\frac12 t\tr B}}{\sqrt{\det \Phi_t}}\,\exp\biggl\{\int_0^t \biggl(
(\Phi_{\tau}^{-1} h_{\tau},\overline A\Phi_{\tau}^{-1} h_{\tau})- (\Phi_\tau^{-1} h_{\tau}, \overline h))\,d\tau\biggr\}.
\end{eqnarray}
At least, the factor $1/2$ at quadratic form   in  exponential \eref{eqBe}  is missed and numerical values
of  \eref{eqS2} and \eref{eqBe}
are different.  As a consequence, the  normalization condition $||e^{it\widehat H}|0\rangle||^2=1$ for evolution of the vacuum state is violated if $h\neq 0$ (see section 8), and perhaps this was  the  reason
for physicists to ignore \cite{Be66} and look for alternative theories (see \cite{GaZo04}, \cite{He00}).
Numerical tests of normalization conditions  \eref{eqS}, \eref{eqS2}, \eref{eqBe} are given in
\cite{statphys}.

Recall that  $\tr (X+Y)=\tr X+\tr Y$ and $\tr X Y=\tr Y X$. In order to represent
the integral $\int_0^t\tr\,\overline\rho_\tau A \,d\tau$ as an algebraic expression, we apply the
R.~Feynman formula \cite{Fe51} for the left and right derivatives, whose
traces coincide:
\begin{eqnarray}
\label{Fe}
\fl \dot C^L_t=\biggl(\frac{d}{dt}\,e^{C_t}\biggr)e^{-C_t}=
\lim_{\Delta t\to0}\frac1{\Delta t}
\int_0^1ds\frac{d}{ds}\,e^{s(C_t+\Delta t \dot C_t)} e^{-sC_t}
=\int_0^1ds\, e^{sC_t}\dot C_t e^{-sC_t},\\
\nonumber
\fl \dot C^R_t=\int_0^1ds\, e^{-sC_t}\dot C_t e^{sC_t},\quad
\tr \dot C^R_t=\int_0^1ds\, \tr \bigl(e^{-sC_t}\dot C_t e^{sC_t}\bigr)=\int_0^1ds\, \tr \dot C_t =\tr \dot C_t.
\end{eqnarray}
These equalities prove that $\tr \dot C^L_t=\tr\dot  C^R_t=\tr\dot  C_t$ because $\tr e^{-sC_t}\dot C_t e^{sC_t}=\tr \dot C_t$.

Consider  the  set of  relationships which follow from 
commutativity of the group $e^{Gt}$ and its generator $G$.
Equations \eref{coeffic} imply explicit algebraic representations of
the left and the right derivatives $C_t^L$ and $C_t^R$:
\begin{eqnarray}
\fl
\label{Feynman}
\dot C_t^L=
\biggl(\frac{d}{dt}\,
\Phi_t^{-1}\biggr)\Phi_t=
-\Phi_t^{-1}\dot\Phi_t= -\Phi_t^{-1}(\Psi_t \overline A-i\Phi_t B)=iB-R_t\overline A,
\\
\nonumber
\fl \dot C_t^R=\Phi_t
\biggl(\frac{d}{dt}\,
\Phi_t^{-1}\biggr)=
-\dot\Phi_t\Phi_t^{-1}= -(A\overline\Psi_t -i B\Phi_t)\Phi_t^{-1}=iB-A\overline \rho_t.
\end{eqnarray}
Since
$\tr C^L_t=\tr C^R_t=\tr C_t$, $e^{\tr C_t}=({\rm
det\,}e^{C_t})=({\rm det\,}\Phi_t)^{-1}$,   we obtain an algebraic
value for the integral of the last summands in \eref{eqS}-\eref{eqS2}
%, which is true for  small	 $t$:
\begin{eqnarray}
\label{part1}
\fl
\nonumber
\int_0^t \tr \,\overline \rho_\tau A\,d\tau = \int_0^t \tr \,R_\tau \overline A\,d\tau =\tr (iBt- C_t),\\
\fl
e^{-\int_0^t
\frac12\,\tr\,\overline\rho_\tau A\,d\tau}=e^{-\frac{1}{2}\tr (iBt- C_t)}=\frac{e^{-\frac{it}{2}\tr B}}{(\pm)\sqrt{\det \Phi_t}},
\\
\label{part2}
\fl
e^{i\widehat H t}\vert_{h=0}=e^{i\widehat H_2 t}=\frac{e^{-\frac{it}{2}\tr B}}{(\pm)\sqrt{\det \Phi_t}}\,e^{-\frac{1}{2} (\ak,R_t\ak)}\,:e^{ (\ak,(\Phi_t^{-1} -I)a)}:\, e^{\frac{1}{2} (a,\overline\rho_t a)}
\end{eqnarray}
with   correctly chosen sign  ($\pm$) which implies the continuity  of  expressions \eref{part1}-\eref{part2} in $t$. If the values of  $e^{\frac12\int_0^t
\tr\,\overline\rho_\tau A\,d\tau-\frac{it}{2}\tr B}$ are not calculated  at a given instant of time $t$, the local choice of the corresponding  branch of the root $(\pm) \sqrt{\det \Phi_t}$ is impossible\footnote{Concerning the proper choice of the sign in  \eref{part1}-\eref{part2},  F.~Berezin  wrote: "It is impossible to remove the remaining non uniqueness in sign." (see \cite{Be66},  p.~136, line 5 from the bottom.)}.

In order to ensure the continuity of functions in \eref{part1}-\eref{part2}, we consider the definition of  the {\it index}.  A similar problem in quasi-classical quantum theory was solved by V.~P.~Maslov  \cite{Ma65},
who introduced the {\it index} for  classical trajectories  as the difference between the number of positive and  negative eigenvalues of
the Hessian matrix of the action along the classical trajectory in configuration space associated to the Hamiltonian,  the set of initial data and independent variables associated to quantum Hamiltonian.

{\small
\begin{figure}[h!]
\begin{minipage}{0.4\linewidth}
\centering\epsfig{figure=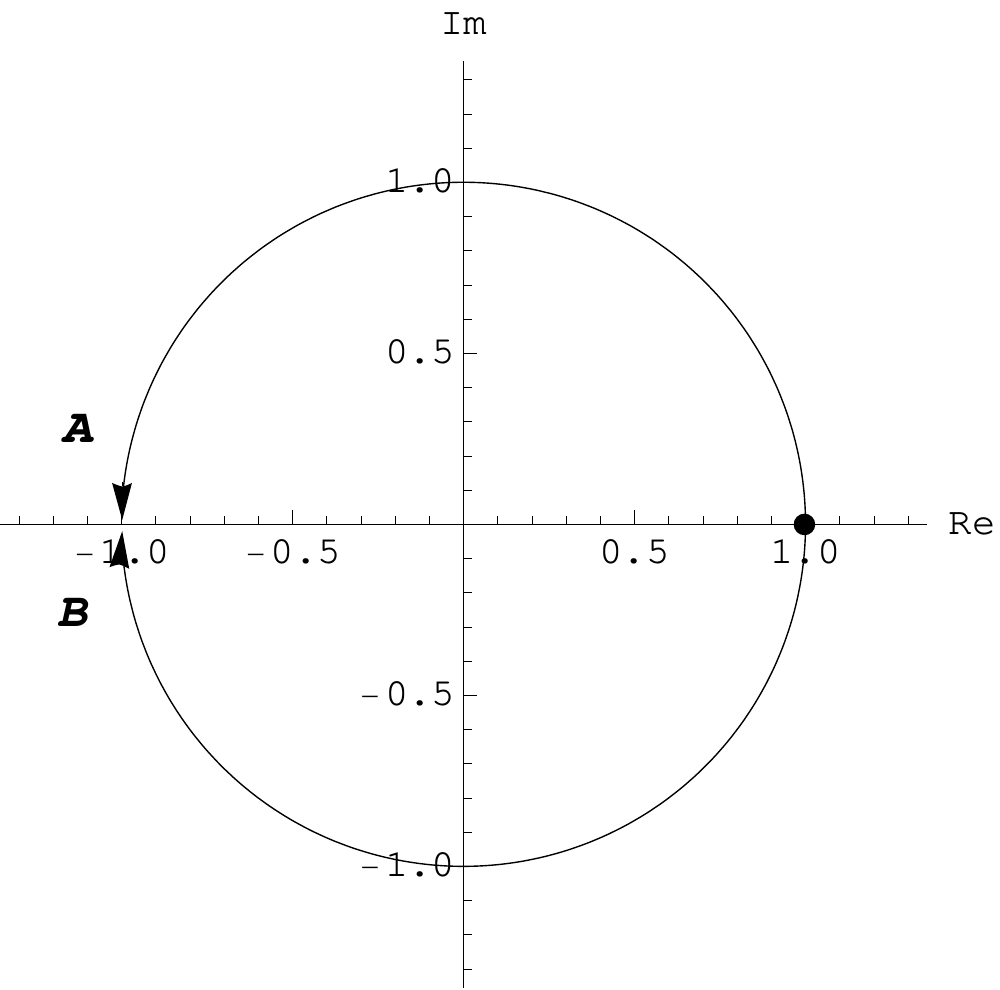,width=\linewidth}
\end{minipage}\hfill
\begin{minipage}{0.5\linewidth}
\centering\epsfig{figure=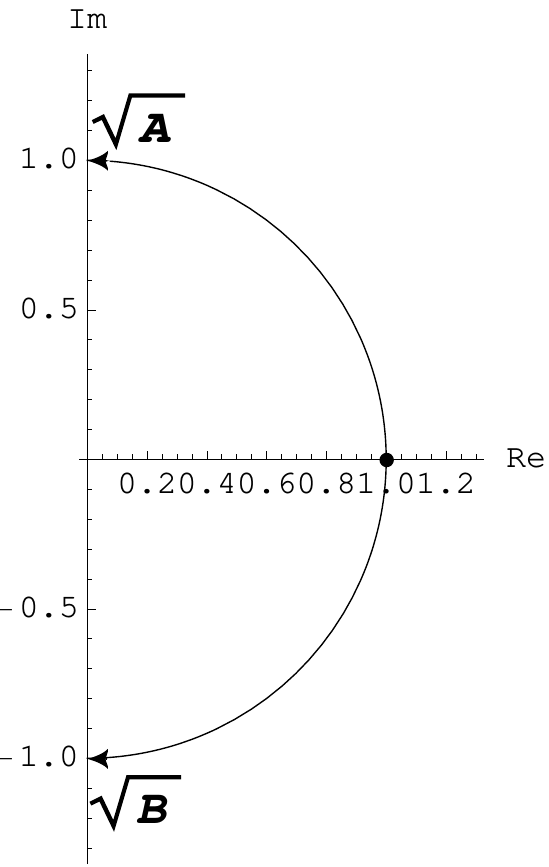,width=0.5\linewidth}
\end{minipage}\hfill
\begin{minipage}{0.4\linewidth}
\centering\epsfig{figure=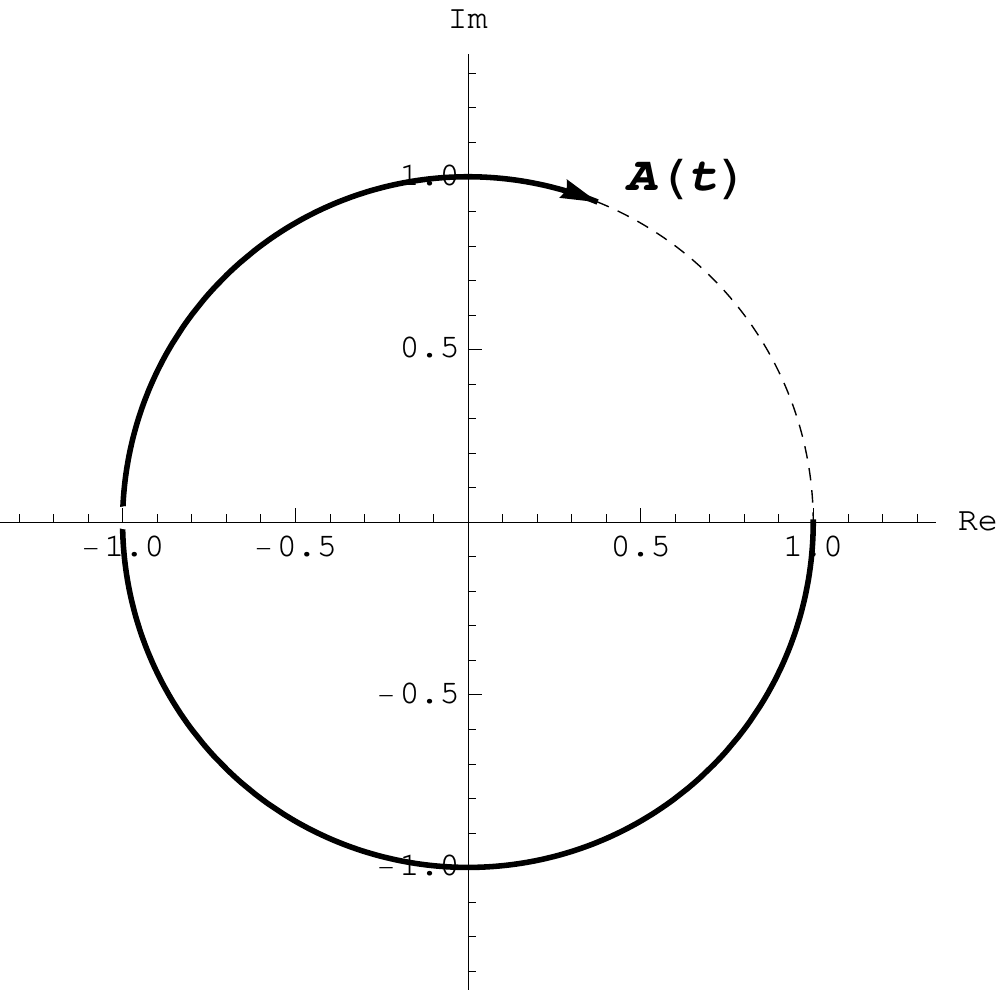,width=\linewidth}
\end{minipage}\hfill
\begin{minipage}{0.4\linewidth}
\centering\epsfig{figure=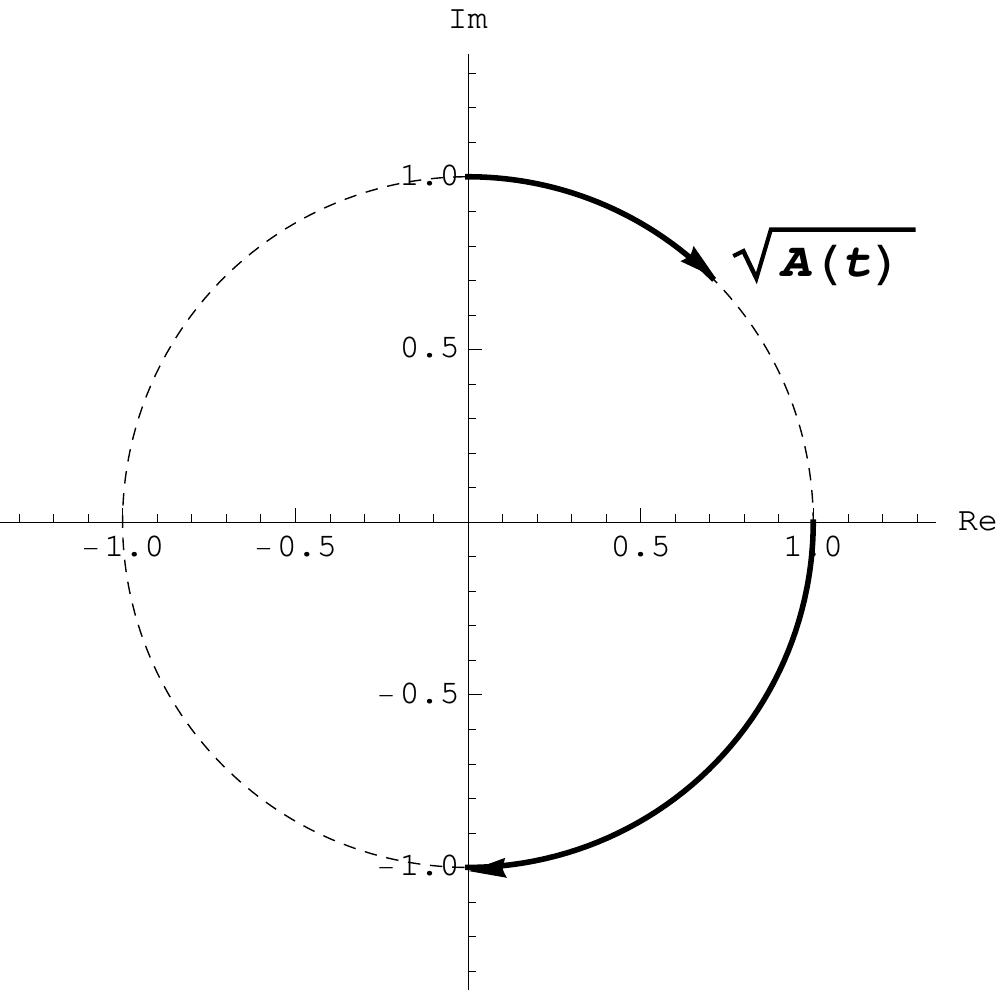,width=\linewidth}
\end{minipage}\hfill
\begin{minipage}{0.33\linewidth}
\centering\epsfig{figure=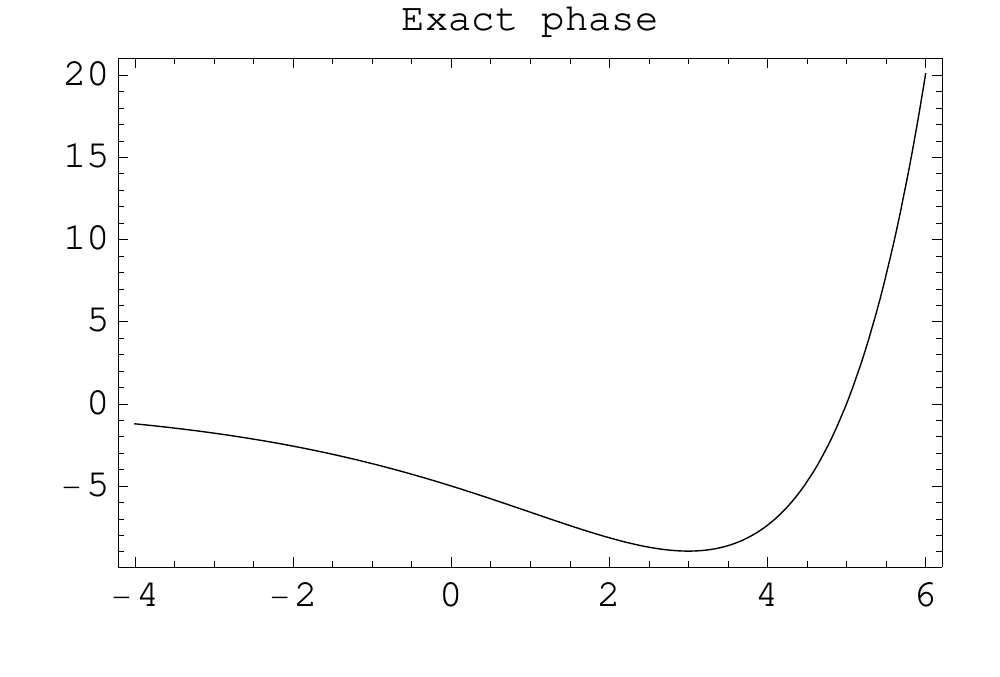,width=\linewidth}
\end{minipage}\hfill
\begin{minipage}{0.33\linewidth}
\centering\epsfig{figure=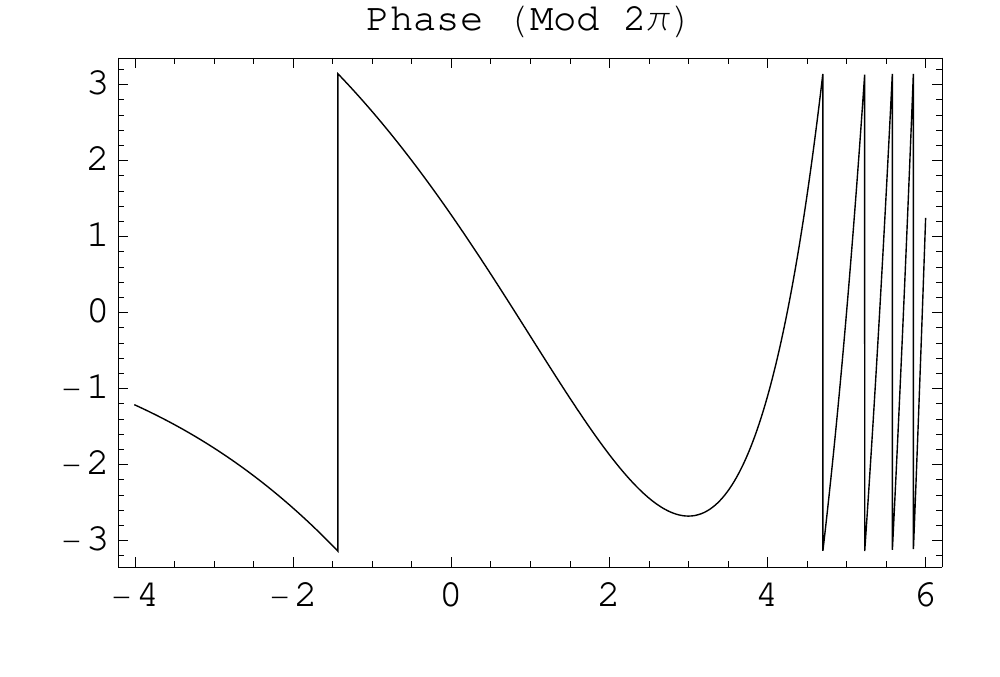,width=\linewidth}
\end{minipage}\hfill
\begin{minipage}{0.33\linewidth}
\centering\epsfig{figure=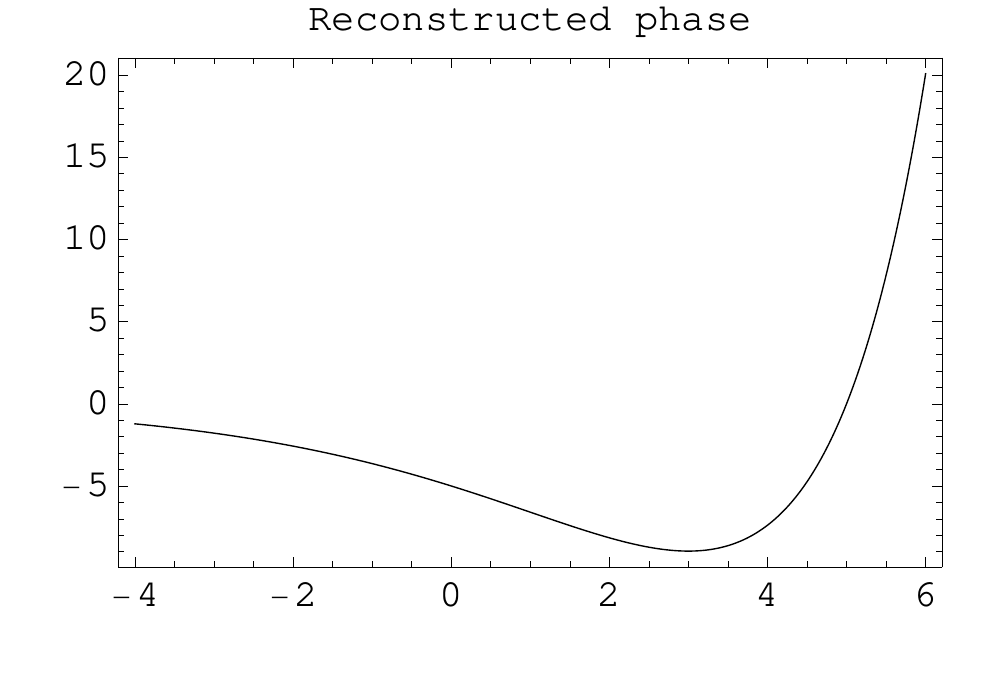,width=\linewidth}
\end{minipage}\hfill
\begin{minipage}{0.33\linewidth}
\centering\epsfig{figure=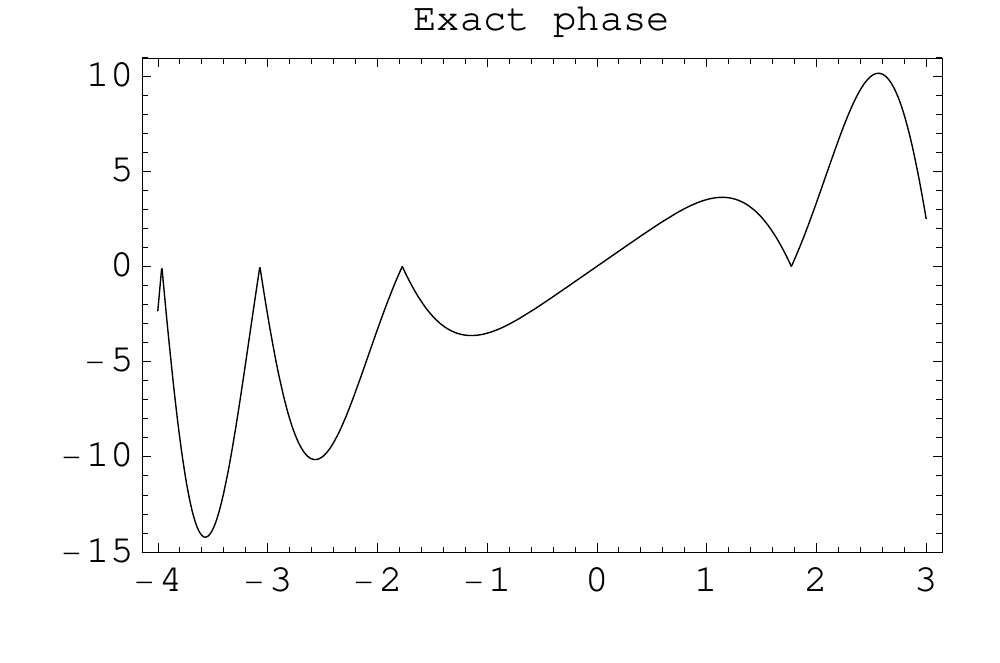,width=\linewidth}
\end{minipage}\hfill
\begin{minipage}{0.33\linewidth}
\centering\epsfig{figure=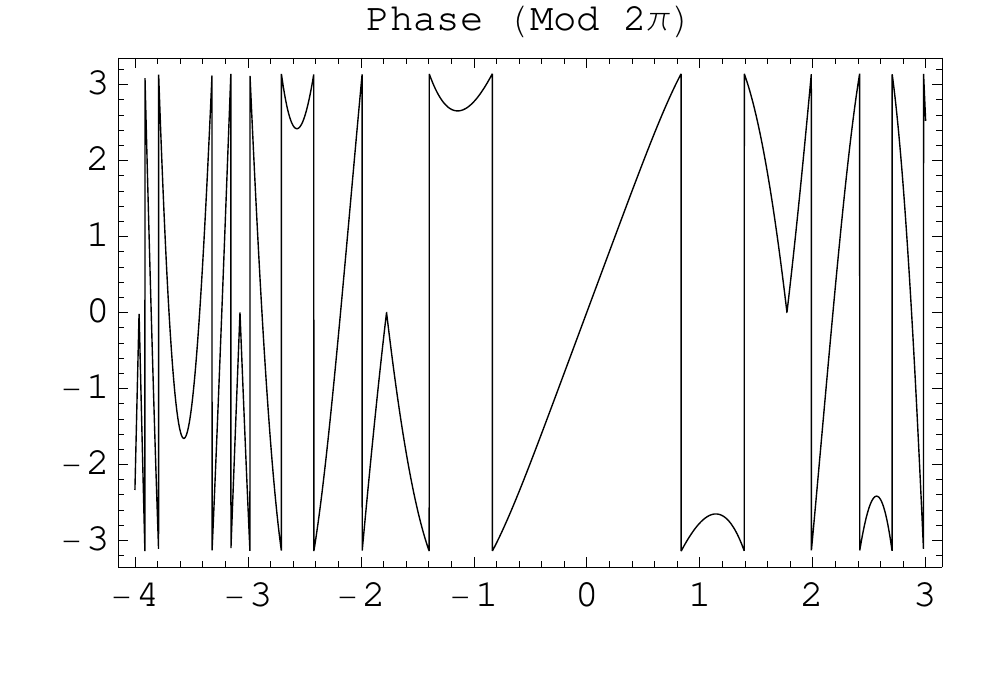,width=\linewidth}
\end{minipage}\hfill
\begin{minipage}{0.33\linewidth}
\centering\epsfig{figure=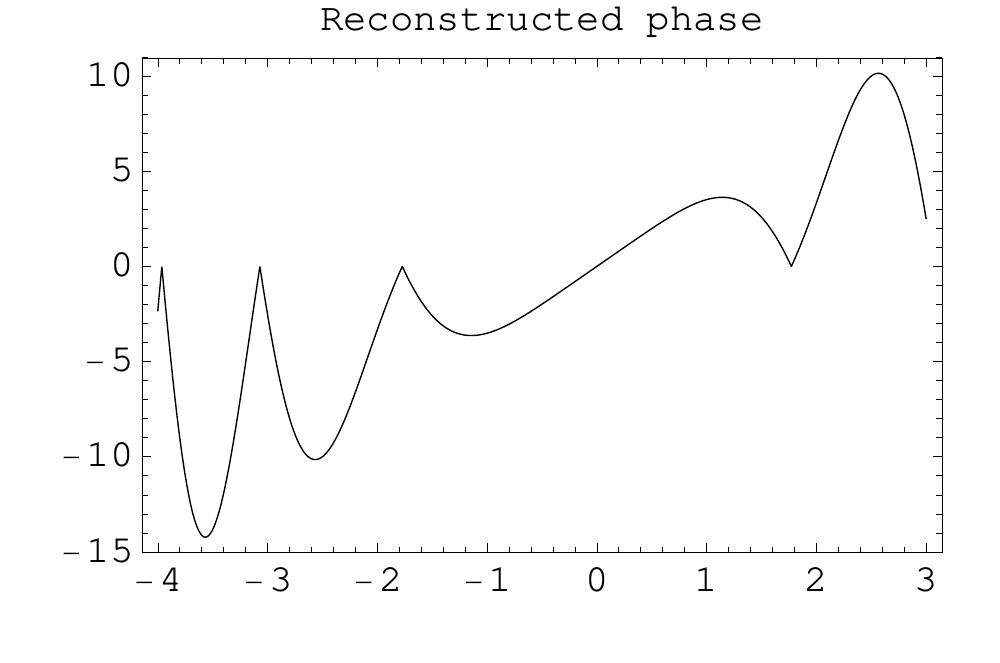,width=\linewidth}
\end{minipage}\hfill
\caption{\footnotesize The first two lines of  figures illustrate
the complex square root calcutated by  MatLab or
Wolfram Mathematica. The complex square root is implemented as a function which is continuous at the positive half line $z>0$ and discontinuous at $z<0$. The last two lines of panels show   successful reconstructions of  the continuous phase functions  based on the $({\rm mod} \,2\pi)$-continuity condition (see equation \eref{defInd} below and the algorithm in \cite{statphys}). }
\end{figure}
}

Recall that numerical values of $\sqrt{z}$  calculated by Wolfram Mathematica (or MatLab, or other modern computational tool) are nonassosiative v.~r.~t. multiplication and discontinuous along the negative cut:
\begin{equation}
\label{ruleSQ}
\fl
\sqrt{z}=\{\sqrt{|z|} e^{\frac{i}{2}\phi}, \;{\rm if} \;\phi\in[0,\pi];\;   \;\sqrt{|z|} e^{\frac{i}{2}(\phi-2\pi)}, \;{\rm if}\;\phi\in(\pi,2\pi)\},
\end{equation}
that is $\phi=\pi$ is the point, where the phase functions are discontinuous  \cite{root}.  At the same time  $\sqrt{\Phi_t}$ in \eref{part1} is continuous
at the origin  and  $\Phi_0=I$.
In the general case,  $\sqrt{{\rm Arg}\det \Phi_t}$ is a  discontinuous function (see the second line of  panels in fig.~1 and the first line of panels in fig.~2), because its  values must belong to half a circle (for example, either to $[0,\pi]$, or to $[-\pi/2,\pi/2]$), and  at each given instant of time $t$ we have no physical or mathematical reasons to choose either positive or negative branch of the root.

On the other hand, the matrix elements  of
{\small $e^{Gt}=\left(
\begin{array}{cc}
                 \Phi_t&\Psi_t\\
                \overline{\Psi}_t&\overline{\Phi}_t
\end{array}
\right)$}
and the left hand sides in \eref{part1}-\eref{part2} are continuous in $t$ (see the second and the third lines
of panels in fig.~2).  The only disadvantage of  representations  \eref{part1}-\eref{part2}  is that
the numerical integration of $\tr \overline \rho_\tau A$ converges very slow  for multimode systems ($n>3$).  Therefore,
the continuity and smoothness of \eref{part2} can be ensured
either by using non-local integral representation \eref{part1} of $s_t$, or by a global continuity construction based on the integer valued {\it index}.

\begin{figure}
\begin{minipage}{0.38\linewidth}
\centering\epsfig{figure=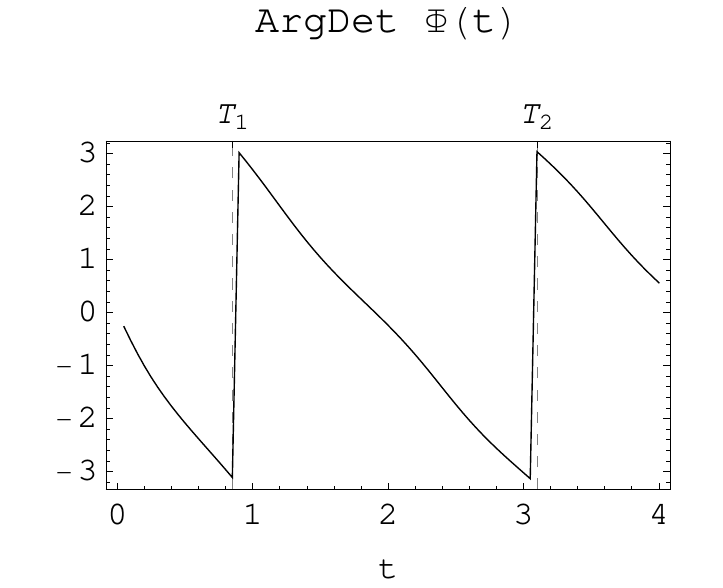,width=\linewidth}
\end{minipage}\hfill
\begin{minipage}{0.38\linewidth}
\centering\epsfig{figure=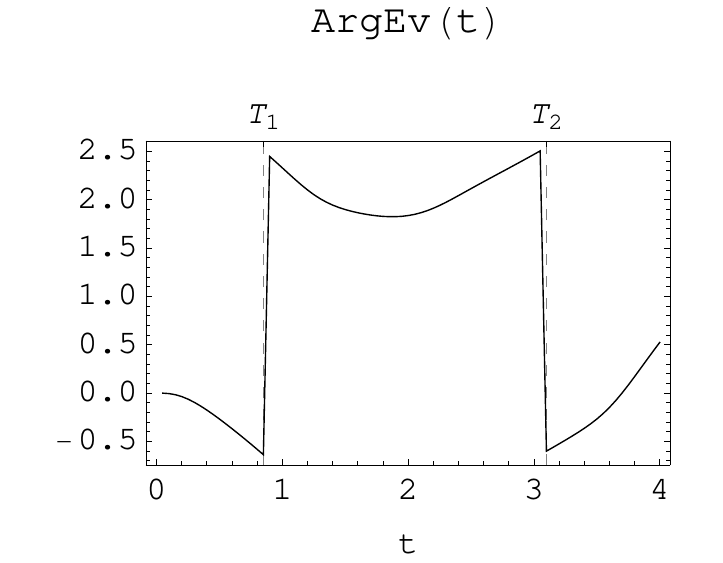,width=\linewidth}
\end{minipage}\hfill
\begin{minipage}{0.38\linewidth}
\centering\epsfig{figure=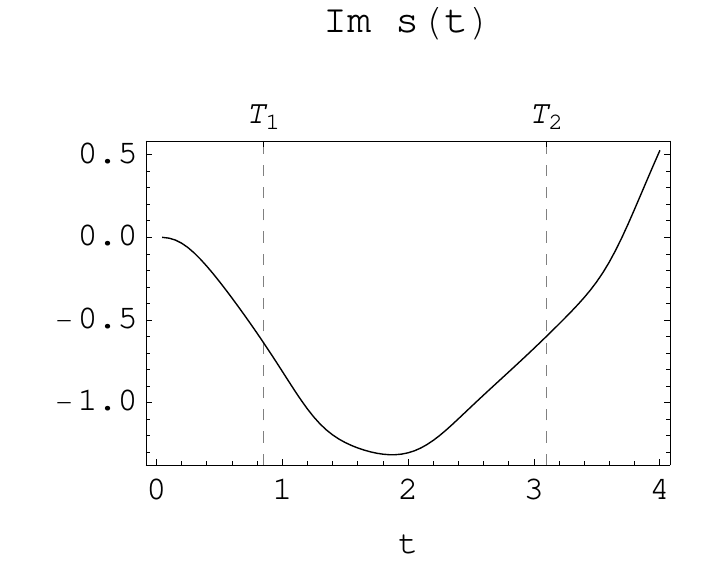,width=\linewidth}
\end{minipage}\hfill
\begin{minipage}{0.38\linewidth}
\centering\epsfig{figure=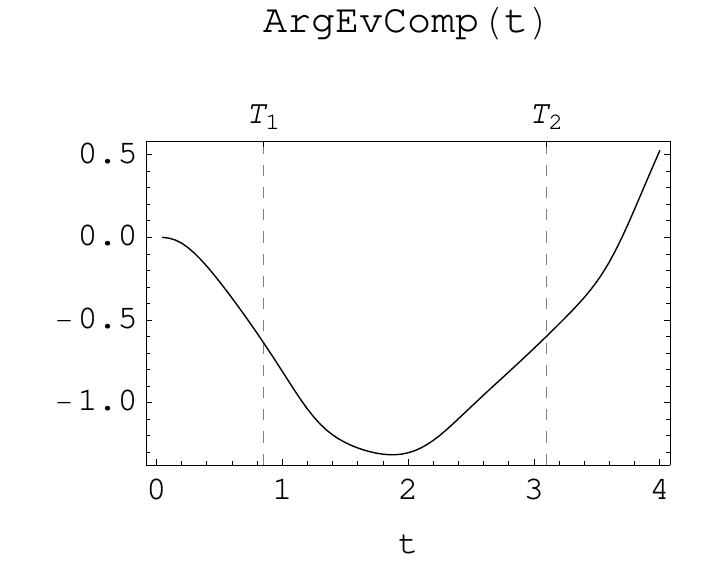,width=\linewidth}
\end{minipage}\hfill
\begin{minipage}{0.38\linewidth}
\centering\epsfig{figure=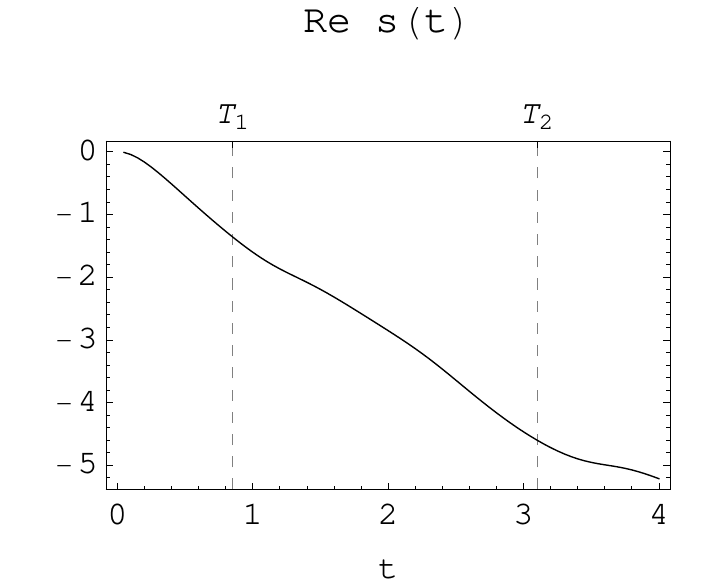,width=\linewidth}
\end{minipage}\hfill
\begin{minipage}{0.38\linewidth}
\centering\epsfig{figure=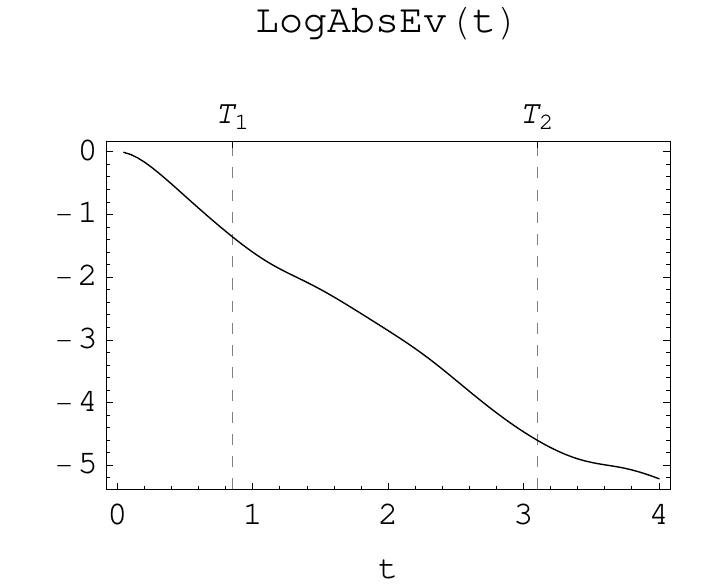,width=\linewidth}
\end{minipage}\hfill
\begin{minipage}{0.38\linewidth}
\centering\epsfig{figure=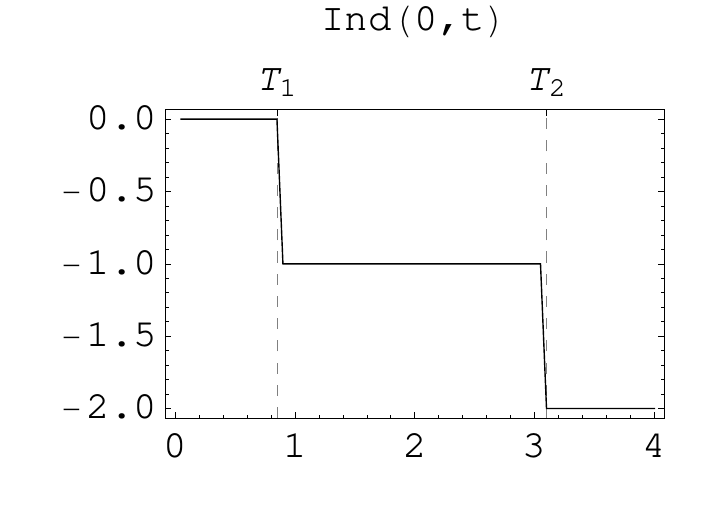,width=\linewidth}
\end{minipage}\hfill
\begin{minipage}{0.38\linewidth}
\centering\epsfig{figure=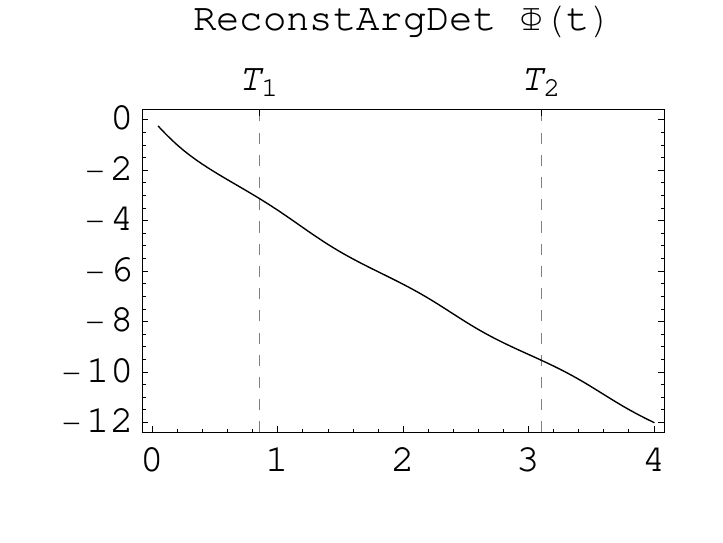,width=\linewidth}
\end{minipage}\hfill
\caption{\footnotesize
The first two panels show the discontinuous  functions
${\rm Arg Det}\,\Phi_t\in [-\pi,\pi]$ and ${\rm Arg\,}\frac{e^{-\frac{it}{2}\tr B}}{\sqrt{\det\Phi_t}}\in [-\pi,\pi]$
in \eref{part1}. The the second line shows continuous functions
${\rm Im}\,s_t\in \R$ (see Eqs. ~\eref{eqS}, \eref{eqS2}) and
${\rm Im }\,S_t$, where
$S_t={\rm Arg }\frac{e^{-\frac{it}{2}\tr B+i\pi\,{\rm Ind}(0,t)}}{\sqrt{\det\Phi_t}}$  differs from the corresponding picture in the
first line  by $i\pi\,{\rm Ind}(0,t)$ in the exponential.
The third  line illustrates coincidence and
continuity of the real part of $\int_0^t
\frac12\,\tr\,\overline\rho_\tau A\,d\tau$ in  Eqs. ~\eref{eqS}  and
${\rm Log}|\frac{e^{-\frac{it}{2}\tr B}}{\sqrt{\det\Phi_t}}|$
in the right hand side of \eref{part1}.
The last two panels show  ${\rm Ind}(0,t)$ defined by \eref{defInd} and reconstructed continuity of  the ${\rm ArgDet} \Phi(t)$
(c.~f.~panel 1 in this figure. }
\end{figure}

An analogue of  the construction of index introduced by V.~P.~Maslov for non-degenerate (recall that $\det |\Phi_t|\ge 1$) and non-Herimitian matrices $\Phi_t$ can be formulated in terms of the polar decomposition $\Phi_t=U_t\,|\Phi_t|$.   The index can be defined correctly, if the arguments of unitary eigenvalues of $U_s$  have a finite number of jumps in a finite time interval $(0, t)$.

Set $\varphi(t)=\frac12\sum_k \lambda_k(t)\in (-\pi,\pi]$, where   $\lambda_k(t)$  are the arguments of eigenvalues $e^{i\lambda_k(t)}$ of $U_t$. Let $\{T_k\}:\;0<T_1<T_2<\dots <T_{n(t)}< t$
be the set of instants of  $2\pi$-jumps of $\varphi(t)$ from one side of the interval $(-\pi,\pi]$ to another during time $t$ (see fig.~2, panel 1). If  $\varphi(t)$ decreases, its jumps from $-\pi$ to $\pi$  are positive, and if $\varphi(t)$ increases, the jumps of the argument are negative (see fig.~1, panels 2 and 5, fig.~2, panels 2 and 5).  Then

\begin{equation}
\label{defInd}
\fl
{\rm Ind} (s,t)\defn -\sum_{T_n\in(s,t)} {\rm sign} (\varphi(T_n+0)-\varphi(T_n-0)) ,
\end{equation}

\begin{equation}
\label{incInd}
\fl
e^{i\widehat H_2 t}=\frac{e^{-\frac{it}{2}\tr B+i\pi{\rm Ind}(0,t)}}{\sqrt{\det \Phi_t}}\,e^{-\frac{1}{2} (\ak,R_t\ak)}\,:e^{ (\ak,(\Phi_t^{-1} -I)a)}:\, e^{\frac{1}{2} (a,\overline\rho_t a)}
\end{equation}
is a continuous function, where the square root and the index are calculated according to \eref{ruleSQ} and
\eref{defInd} respectively.   Examples of continuous reconstructions of the phase functions are shown in the last two lines of panels in fig.~1.

In next section, we derive a pure algebraic representation of $s_t$ for fast numerical implementation.

\section{Algebraic forms of  $e^{s_t}$ and normal symbols of squeezings}
For $z\in \mathbb C^n$, define the normalized  coherent vector   $\psi(z)=e^{(z,\ak)-(\overline
z,a)}|0\rangle=|z\rangle$. By \eref{part2}
and by definition of the normal ordering, the normal symbol of
$e^{i\widehat H_2 t}$ is equal to
\begin{equation*}
\fl
\langle z|e^{i\widehat H_2 t}|z\rangle=\frac{e^{i {\rm Ind}(0,t) -\frac{it}{2}\tr B}}{\sqrt{\det \Phi_t}}\,e^{-\frac{1}{2} (\overline z,R_t \overline z)}\,
e^{ (\overline z,\bigl(\Phi_t^{-1}-I\bigr) z)}\, e^{\frac{1}{2} (z,\overline\rho_t z)},
\end{equation*}
and the commutation rule $e^{-(\ak,z)+(a,\overline z)}F(\ak,a)
e^{(\ak,z)-(a,\overline z)}=F(\ak+\overline z,a+z)$ implies that
\begin{eqnarray*}
\fl
\frac{e^{i {\rm Ind}(0,t) -\frac{it}{2}\tr B}}{\sqrt{\det \Phi_t}}\,e^{-\frac{1}{2} (\overline z,R_t \overline z)}\,e^{ (\overline z,\bigl(\Phi^{-1}_t-I\bigr) z)}\,
 e^{\frac{1}{2} (z,\overline\rho_t z)}=\langle z|e^{i\widehat H_2 t}|z\rangle
=\langle 0|e^{iH_2(\ak+\overline z,a+z) t}|0\rangle
\\
\fl
=\langle 0|e^{\bigl(i\widehat H_2-(\ak,x)+(a,\overline x)\bigr)t}|0\rangle\,
e^{it (\overline z,Bz)-\frac12(\overline z,A\overline z) +\frac12(z,\overline A z) }=e^{s_t}\,e^{it{\rm Im\,}(z,\overline h) 	 },
\nonumber
\end{eqnarray*}
where we set $x=A\overline z-iBz$ and $\overline x=\overline Az+i\overline B\overline z$.
If $\det G\neq 0$, the equations  $A\overline
z-iBz=h$, $\overline Az+i\overline B \overline z= \overline h$
are solvable with respect to $\{z,\overline z\}$ so that
{\small
\begin{equation}
\label{z}
\fl
\left(\begin{array}{c}
                 z(h,\overline h)\\
                 \overline z(h,\overline h)
 \end{array} \right)=
G^{-1}
 \left(\begin{array}{c}
                 h\\
                 \overline h
 \end{array} \right),\quad
\widehat H_2+i(\ak,(A\overline z-iBz))-i(a,(\overline Az+i\overline B\overline z))\vert_{z(h,\overline h)}
= \widehat H.
\end{equation}
} \noindent

Taking into account \eref{Ut}, \eref{part1}, and
\eref{z},  under  assumption ${\rm det}\, G\neq 0$,  we obtain an algebraic expression for the scalar  multiplier $e^{s_t}$:
\begin{eqnarray}
\fl
\label{ScalarFactor}
e^{s_t}=\langle 0|e^{i\widehat Ht}|0\rangle=
\frac{e^{-\frac{it}{2}\tr B+Q_t}}{\sqrt{\det \Phi_t}},\\
\fl
Q_t=i{\rm Ind}(0,t)+
\frac{1}{2} (z,(\overline\rho_t-\overline A t) z)-\frac{1}{2} (\overline z,(R_t-At) \overline z)+(\overline z,(\Phi^{-1}_t-I-iBt) z)\vert_{z(h,\overline h)}
\nonumber.
\end{eqnarray}
Note that the second exponential can be represented  as a symmetric quadratic form in terms of algebraic operations:
\begin{eqnarray*}
\fl
Q_t=\frac12\left(
G^{-1}\left(
\begin{array}{c}
                 h\\
                 \overline h
 \end{array}
\right),
\left(
\begin{array}{cc}
                 \overline\rho_t -\overline A t& (\Phi_t^T)^{-1}-I-i\,\overline Bt\\
                (\Phi_t)^{-1}-I-i Bt& -R_t+At
\end{array}
\right)
G^{-1}\left(
\begin{array}{c}
                 h\\
                 \overline h
 \end{array}
\right)\right).
\end{eqnarray*}

The final result of this section is an  algebraic representation of $s_t$ in terms of the matrices $e^{Gt}$, $G^{-1}(e^{Gt}-I)$, and $G^{-2}(e^{Gt}-I-Gt)$ which are well defined for degenerate and nondegenerate matrices $G$.

For    $\widehat H=\widehat H_2-(\ak,h)+(a,\overline h)$,
we have $e^{s_t}=\langle 0|e^{i\widehat H t}|0\rangle=\langle \phi_t|e^{-i\widehat H_2t}e^{i\widehat H t}|0\rangle$
with
$\phi_t\defn e^{-i\widehat H_2t}|0\rangle\in \otimes_1^n\ell_2$.
By unitary  isomorphism between $\otimes_1^n\ell_2$ and
${\mathcal L}_2({\mathbb R}^n)$
$$
\otimes_1^n\ell_2\ni |0\rangle\leftrightarrow \frac{e^{-\frac{1}{2}|x|^2}}{\pi^\frac{n}{4}}\in {\mathcal L}_2({\mathbb R}^n),
\quad
a\leftrightarrow \frac{x+\partial_x}{\sqrt{2}},
\quad a^\dagger\leftrightarrow\frac{x-\partial_x}{\sqrt 2},
$$
decomposition \eref{genDec} implies
\begin{eqnarray}
\nonumber
\fl
e^{-i\widehat H_2t}|0\rangle=\frac{e^{i {\rm Ind}(0,t) +\frac{it}2\tr B}}{\sqrt{\det \Phi_{-t} }}\, e^{-\frac12(\ak,R_{-t}\ak)}|0\rangle
\leftrightarrow
\frac{e^{i {\rm Ind}(0,t) +\frac{|x|^2}2-(x,(I-R_{-t})^{-1}x)}}{\pi^{\frac{n}4}\sqrt{\det \Phi_{-t}
\det(I-R_{-t})}}\defn \phi_t(x)
\\
\fl
\label{dec2}
=\frac{e^{i {\rm Ind}(0,t) +\frac{|x|^2}2-(x,(I+\rho_{t})^{-1}x)}}{\pi^{\frac{n}4}\sqrt{\det \Phi_{t}
\det(I+\rho_{t})}}\defn \phi_t(x),
\end{eqnarray}
where  $ \phi_t(x)\in {\mathcal L}_2({\mathbb R}^n)$, and $R_{-t}=-\rho_t$, $\Phi_{-t}=\Phi^*_t$.
Moreover, the CCR \eref{CR} implies two  useful identities for determinants:
$\det\Phi_t\,\det\overline \Phi_t\,\det(I-R_t\overline{R_t})=1$ and
{\small
\begin{eqnarray}
\label{equal2}
\fl
\det\Phi_t\,\det\overline \Phi_t\,\det(I-R_t)\,\det(I-\overline{R_t})\det\biggl((I-R_t)^{-1}+(I-\overline R_t)^{-1}-I\biggr)=1.
\end{eqnarray}
}

Let us prove the unitary equivalence of   exponential vectors from $\ell_2$ and ${\mathcal L}_2$:
\begin{eqnarray}
\fl
\nonumber
\psi_t=e^{-i\widehat H_2t}e^{i(\widehat H_2-(\ak,h)+(a,\overline h))t}|0\rangle =e^{-(\ak,h_t)+
i\gamma_t-\frac{|h_t|^2}2}|0\rangle\leftrightarrow
\\
\label{dec1}
\leftrightarrow
\frac{e^{i\gamma_t-\frac{(h_t,\overline{h}_t-h_t) }{2}}}{\pi^{\frac{n}4}}e^{-\frac12(x+\sqrt 2 h_t,x+\sqrt 2 h_t)}
\defn \psi_t(x),
\end{eqnarray}
where $h_t$, $\overline h_t$ are the same as in \eref{sys1}, and
\begin{eqnarray}
\label{gamma}
\fl
\gamma_t=\im \int_0^t(\dot h_s,{\overline h_s})ds,\;
e^{s_t} =\langle\phi_t,\psi_t\rangle_{l_2}=\int_{{\mathbb R}^n}\overline\phi_t(x)\psi_t(x)d^nx
=\int_{{\mathbb R}^n}\phi_{-t}(x)\psi_t(x)d^nx.
\end{eqnarray}
By taking the time  derivative of the left hand side of \eref{dec1} in $\ell_2$ representation, we obtain
{\small
\begin{eqnarray*}
%\label{gamma}
\fl
(\Phi_{-t}a+\Psi_{-t}\ak,\overline h)-(\overline \Phi_{-t}\ak+\overline \Psi_{-t}a,h)-
(a+h_t,\dot{\overline h_t})+(\ak,\dot h_t)+i\dot\gamma_t-\frac{(h_t,\dot{\overline h_t})-(\dot h_t,{\overline h_t})}2=0.
\end{eqnarray*}
}
\noindent
Note that zero values of coefficients at $a$, $\ak$, and  $I$ are the necessary conditions for this equality. Taking into account the  identities
$\Phi_{-t}=\Phi_t^*$, $\Psi_{-t}=-\Psi_t^T$, we obtain the following equations
\begin{eqnarray*}
%\label{gamma}
\fl
\left(\begin{array}{c}
                 \dot h_t\\
                 \dot{\overline h_t}
\end{array}\right)=
\left(\begin{array}{cc}
                 \Phi_t&\Psi_t\\
                 \overline \Psi_t&\overline\Phi_t
\end{array}\right)
\left(\begin{array}{c}
                 h_t\\
                 \overline h_t
\end{array}\right),\quad
i\dot\gamma_t=\frac{(\dot h_t,\overline h_t)-
(h_t,\dot {\overline h_t})}2=i \im (\dot h_t,\overline h_t).
\end{eqnarray*}
Consider the integral representation of \eref{gamma}
\begin{eqnarray*}
%\label{gamma}
\fl
\left(\begin{array}{c}
                 h_t\\
                 {\overline h_t}
\end{array}\right)=\frac{e^{Gt}-I}{G}\left(\begin{array}{c}
                 h\\
                 {\overline h}\end{array}\right),
      \quad
i\gamma_t=i\im \int_0^t(\dot h_s,\overline h_s)ds= \frac12 \int_0^t\det \left(\begin{array}{cc}
    \dot h_t&{\overline {\dot h}}_t\\
                 h_t&\overline h_t
           \end{array}\right)ds
\end{eqnarray*}
and let us transform the above integral  to algebraic form:
{\small
\begin{eqnarray}
\label{gammaT}
\fl
i\gamma_t=\frac12\left(\frac{I-Gt-e^{-Gt}}{G^{2}}\,
\left(
\begin{array}{c}
                 h\\
                 \overline h
 \end{array}
\right),
\left(
\begin{array}{c}
                \overline h\\
                  -h
 \end{array}
\right)
\right),\quad e^{Gt}\equiv \left(\begin{array}{cc}
                 \Phi_t&\Psi_t\\
                 \overline \Psi_t&\overline\Phi_t
\end{array}\right).
\end{eqnarray}
}
The symplectic property of canonical transformations \eref{inversion} implies
{\small
\begin{eqnarray*}
\fl
 \left(\begin{array}{cc}
                 \Phi_t&\Psi_t\\
                 \overline \Psi_t&\overline\Phi_t
\end{array}\right)^T
\,
\left(
\begin{array}{cc}
             0&I\\
-I&0
 \end{array}
\right)\,
 \left(\begin{array}{cc}
                 \Phi_t&\Psi_t\\
                 \overline \Psi_t&\overline\Phi_t
\end{array}\right)
=\left(
\begin{array}{cc}
             \Phi_t^T&\Psi_t^*\\
\Psi_t^T&\Phi_t^*
 \end{array}
\right)
\,
 \left(\begin{array}{cc}
                 \overline\Psi_t&\overline\Phi_t\\
                 -\Phi_t&-\Psi_t
\end{array}\right)
=
 \left(\begin{array}{cc}
                 0&I\\
                 -I&0
\end{array}\right),
\\
\fl
\left(
\begin{array}{cc}
             0&I\\
-I&0
 \end{array}
\right)\,e^{Gt}=
\left(
\begin{array}{cc}
             0&I\\
-I&0
 \end{array}
\right)\,
 \left(\begin{array}{cc}
                 \Phi_t&\Psi_t\\
                 \overline \Psi_t&\overline\Phi_t
\end{array}\right)
=
 \left(\begin{array}{cc}
                 \Phi_{-t}^T&\Psi_{-t}^*\\
                 \Psi_{-t}^T&\Phi_{-t}^*
\end{array}\right)
\, \left(\begin{array}{cc}
                 0&I\\
                 -I&0
\end{array}\right)
=
e^{-G^Tt}\, \left(\begin{array}{cc}
                 0&I\\
                 -I&0
\end{array}\right).
\end{eqnarray*}
}
Hence from equation \eref{sys1}  we have
{\small
\begin{eqnarray*}
\fl
2\,\im (h_s,\dot{\overline h_s})=\det
\left(
\begin{array}{cc}
              h_s&\dot h_s\\
                  \overline h_s& \dot{\overline h_s}
 \end{array}
\right)=
\left(\frac{e^{Gt}-I}{G}\,
\left(
\begin{array}{c}
                 h\\
                  \overline h
 \end{array}
\right),
\,
\left(
\begin{array}{cc}
             0&I\\
-I&0
 \end{array}
\right)\,
e^{Gt}
\left(
\begin{array}{c}
                h\\
                \overline h
 \end{array}
\right)
\right)=
\\
\fl
\left(\frac{e^{Gt}-I}{G}\,
\left(
\begin{array}{c}
                 h\\
                  \overline h
 \end{array}
\right),
\,
e^{-G^Tt}
\left(
\begin{array}{cc}
             0&I\\
-I&0
 \end{array}
\right)\,
\left(
\begin{array}{c}
                h\\
                \overline h
 \end{array}
\right)
\right)
=
\left(\frac{I-e^{-Gt}}{G}\,
\left(
\begin{array}{c}
                 h\\
                  \overline h
 \end{array}
\right),
\,
\left(
\begin{array}{c}
                \overline h\\
                -h
 \end{array}
\right)
\right).
\end{eqnarray*}
}
\noindent
Integration of this equality in $s$ over $[0,t]$ readily implies \eref{gammaT}.
Finally, by combining \eref{dec2}  and \eref{dec1}, we obtain an algebraic expression  for $e^{s_t}=\langle0|e^{i\widehat H_t}|0\rangle$ and also expressions for
$\psi_z=e^{S_t-\frac12(\ak,R_t\ak)-(G_t,\ak)}|z\rangle$ and $N_{A,B,h}(\overline z,z)=\langle z|\psi_z\rangle$ as corollaries.

\begin{thm} {\bf 2}.
{\it
\begin{itemize}
\item[1.] For abitrary symmetric matrix $A$, Hermitian matrix $B$, and complex vector $h$, the vacuum expectation of the unitary group $e^{it\widehat H}$ \eref{Ut} is equal to
\begin{eqnarray}
\fl
\nonumber
e^{s_t}=\langle 0|e^{it\widehat H}|0\rangle=e^{i {\rm Ind}(0,t) +i\gamma_t-\frac12\,(h_t,h_t)-\frac12\,(h_t,(\overline h_t- h_t))}\int d^nx\frac{e^{-\sqrt2(h_t,x)-(x,(I+\rho_t^*)^{-1}x)}}{\pi^{n/4}
\sqrt{\det(I+\rho^*_t)\det \Phi_t}}\\
\fl
\label{vac}
=\frac{e^{i {\rm Ind}(0,t) + i\gamma_t-\frac{1}{2}((h_t,(\overline h_t-\overline \rho_th_t))+it\tr B)}}{\sqrt{\det \Phi_t}}=\frac{e^{i {\rm Ind}(0,t) +i\gamma_t-\frac{it}{2}\tr B-\frac{1}{2}(\overline h_{-t},\Phi_t^{-1} h_t)}}{\sqrt{\det \Phi_t}},
\end{eqnarray}
where  $\overline \rho_t=\rho_t^*$, $h_t$  and $\gamma_t$ are given by \eref{sys1} and \eref{gammaT}.

\item[2.]  The state $\psi_z=e^{S_t-\frac12(\ak,R_t\ak)-(G_t,\ak)}|z\rangle$
 is a unit vector in
$\otimes_1^n\ell_2,e^{s_t}$
and its image in $\mathcal{L}_2({\mathbb R}^n)$
is equal to the Gaussian function
\begin{eqnarray}
\fl
\label{XState}
\psi_t(x)=\frac{e^{S_t}}{\pi^{\frac{n}4}\sqrt{{\rm det}(I-R_t)}}\,
e^{\frac12 |x|^2-(x+\frac{G_t}{\sqrt2},(I-R_t)^{-1}(x+\frac{G_t}{\sqrt2}))}\in {\mathcal L}_2({\mathbb R}^n),
\end{eqnarray}
where
$G_t=\Phi_t^{-1}(h_t-z)$, $S_t=s_t+(z,\overline f_t-\frac12\,(\overline z-\overline \rho_t z))$, and $f_t=h_t-\rho_t\overline h_t$ (see \eref{solEq}).
\item[3.] The {\it normal symbol of squeezing} \eref{Ut} is equal to
\begin{eqnarray}
\label{normalSymb}
\fl
N_{A,B,h}(\overline z,z)\defn\langle z|e^{i\widehat H t}|z\rangle
=e^{{\rm Ind}_t +s_t-|z|^2-
\frac12(\overline z , R_t\overline z)-(v_t,\overline z)+(\overline z,(\Phi_t^{-1}-I)z)+\frac12(z,\overline \rho_t z)+(\overline f_t,z))}\,.
\end{eqnarray}
\end{itemize}
}
\end{thm}
The coincidence of expressions \eref{eqS}, \eref{eqS2}, \eref{ScalarFactor}, \eref{vac} was tested numerically. The testing  modules are available for users of Wolfram Mathematica   at  \cite{statphys}.

\section{Inner product  of squeezed states and  composition of squeezings }
The inner products of squeezed states are necessary for constructing orthonormal bases, and the symbols of compositions of squeezings allow one to represent in algebraic terms the quantum evolution of multimode systems in some important cases.

In this section we use the well known canonical isometric isomorphysm between $\otimes_1^n\ell_2$ and
${\mathcal L}_2({\mathbb R}^n)$, so that $|0\rangle\leftrightarrow \frac{e^{-\frac{1}{2}x^2}}{\pi^\frac{n}{4}}$ and
$a\leftrightarrow\frac{x+\partial_x}{\sqrt{2}}$,
$a^\dagger\leftrightarrow\frac{x-\partial_x}{\sqrt2}$.
According to  equation (4.1) from \cite{ChRaTl11}, the  multimode squeezed state
\begin{eqnarray*}
\fl
e^{i\widehat H t}|z\rangle= e^{s_t-\frac12(\ak,R_t\ak)-(g_t,\ak)}
|0\rangle\in\otimes_1^n\ell_2,\quad z\in \mathbb C^n
\end{eqnarray*}
is unitary equivalent to the Gaussian $\psi$-function
\begin{eqnarray*}
\fl
%\label{XState}
\psi_t(x)=\frac{e^{s_t}}{\pi^{\frac{n}4}\sqrt{{\rm det}(I-R_t)}}\,
e^{\frac12 |x|^2-(x+\frac{g_t}{\sqrt2},(I-R_t)^{-1}(x+\frac{g_t}{\sqrt2}))}\in {\mathcal L}_2({\mathbb R}^n),
\end{eqnarray*}
where
$g_t=\Phi_t^{-1}h_t$.

The calculation of the norm $||\psi_t||^2_{{\mathcal L}_2}$
reduces to integration of the Gaussian function $\overline\psi_t(x)\psi_t(x)$.
Note that $R_t=R^T_t$, $\rho_t=\rho^T_t$, and \eref{CR} imply a set of useful   identities:
\begin{eqnarray*}
\fl
I-R_tR^*_t=I-R_t\overline R_t=|\Phi_t|^{-2},\quad {\rm det}(I-R_t\overline R_t)\,{\rm det}\Phi_t\,{\rm det}\overline \Phi_t=I,\\
\fl
\Omega_t=(I-\overline R_t)^{-1}+(I-R_t)^{-1}-I
=(I-\overline R_t)^{-1}(I-\overline R_tR_t)(I-R_t)^{-1}=\overline\Omega_t=\Omega^T_t,
\end{eqnarray*}
and $\Omega_t^{-1}=(\Phi^T_t-\Psi_t^T)(\overline\Phi_t- \overline\Psi_t)=
(\Phi_t^*-\Psi_t^*)(\Phi_t- \Psi_t)$. Therefore, $\overline\psi_t(x)\psi_t(x)$ is a well defined Gaussian
density with  correlation matrix $\Omega_t>0$. After integration of a product of Gaussian functions \eref{XState} we obtain
 \begin{eqnarray}
\fl
||\psi_t||_{{\mathcal L}_2}^2=\frac1{\sqrt{\det(I-\overline R_t R_t)}}
e^{2{\rm Re}\,s_t+2({\rm Re\,}(I-R_t)^{-1}g_t,\Omega^{\,-1}{\rm Re\,}(I-R_t)^{-1}g_t)-{\rm Re\,}(g_t,(I-R_t)^{-1}g_t)}
=1
\label{gaussNorm}
\end{eqnarray}
because  from $e^{2{\rm Re}\,s_t} =\sqrt{\det(I-\overline R_t R_t)}$ and
$\det M =\det M^T$ we have
 \begin{eqnarray*}
 \fl
\frac{e^{2{\rm Re}\,s_t}}{\sqrt{\det(I-\overline R_t R_t)}}
=\sqrt{\det{(\Phi_t\Phi_t^*-\Psi_t\Psi_t^*)}}^{-1}=1.
\end{eqnarray*}
On the other hand,
 \begin{eqnarray*}
\fl
{\rm Re\,}(I-R_t)^{-1}g_t,\Omega^{\,-1}{\rm Re\,}(I-R_t)^{-1}g_t)-{\rm Re\,}(g_t,(I-R_t)^{-1}g_t)
=0.
\end{eqnarray*}

%This normalization condition was successfully tested numerically in dimensions up to 6.
% with integral  (i)  and algebraic  (a) representations of $s_t$:  $T_a =$
%$T_(i)\approx 180 sec$ for $n=5$  and   $T_a =$
%$T_(i)\approx 180 sec$ for $n=5$

 Similarly, for
$G_k=\Phi_t^{-1}(h_k-z_k)$, $S_k=s_k+(\overline f_k,z_k)+\frac{(z_k,\overline \rho_k z_k)}2-\frac12|z_k|^2$,
$R_k=\Phi_k^{-1}\Psi_k$ $ (k=1,2)$, and
$Y=(I-\overline R_1)^{-1}\overline G_1+(I-R_2)^{-1} G_2$, we calculate the inner product of squeezed states in ${\mathcal L}_2({\mathbb R}^n)$
or $\otimes_{1}^n\ell_2$ representation:
{\small
\begin{eqnarray}
\label{innerP}
\fl
\langle\psi_1,\psi_2\rangle_{{\mathcal L}_2}=
e^{\overline S_1+S_2}\,
\int
\frac{e^{|x|^2-\bigl((x+\frac1{\sqrt2}\Omega_{12}^{-1}Y),\Omega_{12}
(x+\frac1{\sqrt2}\Omega_{12}^{-1}Y)\bigr)}}{\pi^{\frac{n}2} \sqrt{\det(I-\overline R_1)\det(I-R_2)}}\,d^nx=
\frac{e^{\sigma_{12}}}{\sqrt{\det(I-\overline R_1 R_2)}},
%=\langle z_1|e^{-i\widehat H_1}e^{i\widehat H_1}|z_2\rangle
\\
\nonumber
\fl
\Omega_{12}=\Omega_{12}^T=(I-\overline R_1)^{-1}+(I-R_2)^{-1}-I=(I-\overline R_1)^{-1}
(I-\overline R_1 R_2) (I-R_2)^{-1},
\\
\fl
\label{sigma12}
\sigma_{12}=\overline S_1+S_2-\frac12\bigl((\overline G_1,(I-\overline R_1)^{-1}\overline G_1)
-\frac12\bigl((G_2,(I- R_2)^{-1}G_2)+\frac12(Y,\Omega_{12}Y).
\end{eqnarray}
}

A simple approach to the  composition of squeezings can be given in terms of canonical transformations. Consider   $U_1=e^{-i\widehat H_1}$, $U_2=e^{i\widehat H_2}$ with unit time $t=1$. We skip here
the time dependence because the semigroup property does not hold for the composition $U_1U_2$. The action  of $U_k$ on functions of $\ak,\,a$ can be expressed in terms of $\Phi_{k}$ and $\Psi_{k}$ by \eref{sys1}. Since the scalar operators $U_k$ commute with numerical expressions or matrices with scalar valued coefficients and act just on the creation-annihilation operators, we have
{\small
\begin{eqnarray*}
\fl
\left(\begin{array}{c}
                 a_2\\
                 \ak_2
\end{array}\right)=
U_2U_{1}
\left(\begin{array}{c}
                 a\\
                 \ak
\end{array}\right)U_{1}^*U_2^*
=
\left(\begin{array}{cc}
                 \Phi_{12}&\Psi_{12}\\
                 \overline \Psi_{12}&\overline\Phi_{12}
\end{array}\right)
\left(\begin{array}{c}
                 a\\
                 \ak
\end{array}\right)
+
\left(\begin{array}{c}
                h_{12}\\
                 \overline h_{12}
\end{array}\right),
\\
\fl
\Phi_{12}=\Phi_1\Phi_2+\Psi_1\overline\Psi_2,\;
\Psi_{12}=\Phi_1\Psi_2+\Phi_1\overline \Psi_2,\quad
\nonumber
  h_{12}=\Phi_{12}h_2+\Psi_{12}\overline h_2+h_1.
\end{eqnarray*}
}
 It can be readily proved that  $\Phi_{12}$ and $\Psi_{12}$ possess the CCR property \eref{CR}. Then
\begin{eqnarray}
\label{comp1}
\fl
U_{12}=e^{s_{12}}e^{-\frac{1}{2} (\ak,R_{12}\ak)-(g_{12},\ak)}\,e^{ (\ak,(\Phi_{12}^{-1}-I)a)}\,
e^{\frac{1}{2} (a,\overline\rho_{12} a)+(\overline f_{12},a)},
\end{eqnarray}
where $R_{12}=\Phi_{12}^{-1}\Psi_{12}$,
$ \overline \rho_{12}=\overline\Psi_{12}\Phi^{-1}_{12}$,
 $g_{12}=\Phi_{12}^{-1}h_{12}$, $f_{12}=h_{12}-\overline\rho_{12} \,\overline h_{12}$, and (see \eref{sigma12})
 \begin{equation}
 e^{s_{12}}=\langle 0|e^{-i\widehat H_1t_1}e^{i\widehat H_2t_2}|0\rangle=
 \frac{e^{\sigma_{12}}}{\sqrt{\det(I-\overline R_1 R_2)}}.
  \end{equation}
These collection of parameters describe the normal ordering of the composition of squeezings:
 \begin{equation}
 U_1U_2
=e^{s_{12}}e^{-\frac{1}{2} (\ak,R_{12}\ak)-(g_{12},\ak)}\,e^{ (\ak,C_{12} a)}\,
e^{\frac{1}{2} (a,\overline\rho_{12} a)+(\overline f_{12},a)}.
  \end{equation}

\section{The Jordan decomposition of squeezings}

In the general case, the Jordan decomposition $G=D J D^{-1}$ justifies
a useful representation of $(2n\times 2n)$-matrix $S_t=e^{Gt}=D e^{Jt} D^{-1}$   as the exponent of the  Jordan matrix $J$ with $(n_k\times n_k)$-blocks $J_k$:
{\small
\begin{eqnarray*}
\fl
J_k\defn\left(\begin{array}{ccccc}
                 \lambda_k&1&0&\dots&0\\
                0&\lambda_k &1&\dots&0\\
                 0&0&\ddots&&\vdots\\
                 0&0&0&0&\lambda_k
\end{array}\right)
\rightarrow
e^{J_kt}=
e^{\lambda_k t}\Delta_k,\quad \Delta_k\defn
\left(\begin{array}{ccccc}
                 1&1&\frac1{2!}&\dots& \frac1{(n_k-1)!}\\
                 0&1&1&\dots& \frac1{(n_k-2)!}\\
                 0&0&\ddots&&\vdots\\
                 0&0&0&0&1
\end{array}\right).
\end{eqnarray*}
}

\noindent
The muliplicity $n_k$ of $\lambda_k$ coincides with the rank of $J_k$,
and decomposition of
\begin{eqnarray}
\label{intEGT}
\fl
F^{(1)}(t)=\frac{e^{Gt}-I}G=D\int_0^t e^{sJ}ds\,D^{-1}=D\,\frac{e^{Jt}-I}{J}\,D^{-1}
\end{eqnarray}
is well defined  in regular  and degenerate cases. The Jordan blocks $J_k$
generate triangle matrices $F^{(1)}_k(t)=\frac{e^{J_kt}-I}{J_k}$:
{\small
\begin{eqnarray}
\label{repht}
\fl
J_k^{-1}=\lambda_k^{-1}
\left(\begin{array}{cccc}
                 1& -\lambda_k^{-1}&\dots&(-\lambda_k)^{-n_k+1}\\
                 0& 1&\dots& (-\lambda_k)^{-n_k+2}\\
                 0&0&\ddots&\vdots\\
                 0&0&\dots& 1
\end{array}\right),\quad
\frac{e^{Gt}-I}G=
D\left(\begin{array}{cccc}
                 F_1^{(1)}(t)&0&\dots &0\\
                0&F_2^{(1)}(t) &\dots &0\\
                 0&0&\ddots &\vdots\\
                 0&0&\dots & F_K^{(1)}(t)
\end{array}\right)D^{-1} ,
\nonumber\\
\fl
F^{(1)}_k(t)\biggr\vert_{\lambda_k=0}=\left(\begin{array}{ccccc}
                t&\frac{1}{2!}t^2&\frac{1}{3!}t^3&\dots&\frac{1}{n_k!}t^{n_k}\\
                0&t&\frac{1}{2!}t^2&\dots&\frac{1}{(n_k-1)!}t^{n_k-1}\\
                 0&0&\ddots&&\vdots\\
                 0&0&0&0&t
\end{array}\right),\quad
(F^{(1)}_k(t))_{ij}=\frac{1-e^{\lambda_k t}
\sum_{m=0}^{j-i}\frac{(-\lambda_k t)^m}{m!}}{(-\lambda_k)^{j-i+1}},
\end{eqnarray}
}

\noindent
for $ i\ge j$;  otherwise, $F^{(1)}_k(i,j)=0$. The matrices
 $F^{(1)}_k(t)$ are well defined in the degenerate case because
$(F^{(1)}_k(t))_{ij}\to \frac{t^{j-i+1}}{(j-i+1)!}$ as $\lambda_k\to0$.

The Jordan  decomposition can be also used  for calculation
of  $\frac{e^{Gt}-I-Gt}{G^2}$ because the algebraic form of
$\frac{e^{J_kt}-I-J_kt}{J_k^2}$ is well defined in nondegenerate and degenerate cases:
\begin{eqnarray}
\label{intEGT2}
\fl
F^{(2)}(t)=D\int_0^td\tau\int_0^\tau e^{sJ}ds\,D^{-1}=\frac{e^{Gt}-I-Gt}{G^2},\quad F^{(2)}_k(t)=\frac{e^{J_kt}-I-J_kt}{J^2_k}.
\quad
\end{eqnarray}
Moreover, the following expressions for components
related to Jordan decomposition are satisfied:
{\footnotesize
\begin{eqnarray}
\label{repht2}
\fl
J_k^{2}=
\left(\begin{array}{cccccc}
                 \lambda_k^2& 2\lambda_k&1&0&\dots&0\\
                 0& \lambda_k^{2}&2\lambda_k&1&\dots&0\\
                 0&0&0&0&\ddots&\vdots\\
                 0&0&0&0&\dots& \lambda_k^{2}
\end{array}\right),\quad
J_k^{-2}=
\left(\begin{array}{cccc}
                 \lambda_k^{-2}& -2\lambda_k^{-3}&\dots&+n_k(-\lambda_k)^{-n_k-1}\\
                 0& \lambda_k^{-2}&\dots& +(n_k-1)(-\lambda_k)^{-n_k}\\
                 0&0&\ddots&\vdots\\
                 0&0&\dots& \lambda_k^{-2}
\end{array}\right),\quad
\\
\fl
\nonumber
\frac{e^{Gt}-I-Gt}{G^2}=
D\left(\begin{array}{cccc}
                 F^{(2)}_1(t)&0&\dots &0\\
                0&F^{(2)}_2(t) &\dots &0\\
                 0&0&\ddots &\vdots\\
                 0&0&\dots & F^{(2)}_K(t)
\end{array}\right)D^{-1} ,
\\
\fl
\label{f2}
(F^{(2)}_k(t))_{ij}=\frac{(-1)^{j-i+1}} {\lambda_k^{i-j+2}} \left(j-i+1+\lambda_kt-e^{\lambda_kt} \sum_{m=0}^{j-i}\frac{(j-i+1-m)(-\lambda_kt)^m}{m!}\right),
\\
\fl
\nonumber
\quad
F^{(2)}_k(t)\biggr\vert_{\lambda_k=0}=-t^2\left(
\begin{array}{cccc}
 \frac{1}{2!} & \frac{-t}{3!} & \dots & \frac{(-t)^{k-2}}{k!}\\
 0 & \frac{1}{2!} &  \dots &\frac{(-t)^{k-3}}{(k-1)!}\\
  0 & 0 & \ddots & \vdots\\
 0 & 0 & \dots & \frac{1}{2!}
\end{array}
\right),
\end{eqnarray}
}
\noindent
for $ i\ge j$;  otherwise, $(F^{(2)}_k(t))_{ij}=0$. The triangle  matrices
 $F^{(2)}_k(t)$ are well defined in the degenerate case because
$(F^{(2)}_k(t))_{ij}\to \frac{t^{j-i+1}}{(j-i+1)!}$ as $\lambda_k\to0$.

This observation establishes an algebraic representation for $h(t)$, $\overline h(t)$, $\gamma_t$, and $s_t$
which follows from \eref{sys1} and \eref{repht} with constant matrices $D$ and well-defined triangle matrices $F^{(1)}_k(t)$,
$F^{(2)}_k(t)$. Implementation time for calculation of $F(t)$ according to
\eref{repht}, \eref{repht2} is  faster than by \eref{intEGT}, \eref{intEGT2} (see \cite{statphys}).

\section{An example of normal decomposition}

In this section, we consider  Hamiltonian \eref{ham} such that $G$ is invertible and all matrices in \eref{coeffic} and
\eref{ScalarFactor} can be described explicitly in
terms of $G$ and the spectral expansions of the  Hermitian matrix
$D=A\overline A -B^2$ in any dimension.

Suppose that the matrix $D=A\overline A -B^2$ is not degenerate and
$BA=A\overline B$. Then $\overline B\,\overline A =\overline A B$,
$BA\overline A=A\overline B\,\overline A=A\overline A B$, and
 $\overline AB^2=\overline B\,\overline A B=\overline B^2\,\overline A$. These relationships imply that
 \begin{eqnarray}
\label{funG}
   \fl
G^{2n}=\left(
                \begin{array}{cc}
                 D^n&0\\
                 0&\overline D^n
                \end{array}
      \right),\quad
      G^{2n+1}=G\,\left(
                \begin{array}{cc}
                 D^n&0\\
                 0&\overline D^n
                \end{array}
      \right)=\left(
                \begin{array}{cc}
                 D^n&0\\
                 0&\overline D^n
                \end{array}
                \right)\,G.
\end{eqnarray}
Hence the matrix  {\small $G^{2}= \left(
                \begin{array}{cc}
                A\overline A -B^2&0\\
                 0&\overline A A -\overline B^2
                \end{array}
      \right)$ } does not degenerate. Therefore,  {\small $G
\stackrel{\rm def}{=} \left(\begin{array}{cc}
                 -iB&A\\
                 \overline A&i\overline{B}
\end{array}\right)$} so  does. Moreover,  the matrices {\small $G^{-2}= \left(
                \begin{array}{cc}
                (A\overline A -B^2)^{-1}&0\\
                 0&(\overline A A -\overline B^2)^{-1}
                \end{array}
      \right)$ } and $G^{-\frac12}$ are well defined in terms of the spectral expansion of $D$, and
 {\small
 \begin{eqnarray}
\label{expG}
\fl
\nonumber
e^{Gt}=\left(\begin{array}{cc}
                 \Phi_{t}&\Psi_{t}\\
                 \overline\Psi_{t}&\overline \Phi_{t}
\end{array}\right)=G \,\left(
                \begin{array}{cc}
                 D^{-\frac12}\sinh D^{\frac12}t&0\\
                 0&{\overline D}^{\,-\frac12}\sinh\overline D^{\frac12}t
                \end{array}
                \right)
+\left(
                \begin{array}{cc}
                 \cosh D^{\frac12}t&0\\
                 0&\cosh\overline D^{\frac12}t
                \end{array}
                \right),
                \\  \nonumber
\\
   \fl
G^{-1}=G^{-2} G=
\left(
                \begin{array}{cc}
             -i  D^{-1}B&D^{-1}A\\
                 {\overline D}^{\,-1}\overline A&i{\overline D}^{\,-1}\overline B
                \end{array}
      \right)=
      \left(
                \begin{array}{cc}
             -i  B D^{-1}&A{\overline D}^{\,-1}\\
                 \overline A D^{-1}&i\overline B\,{\overline D}^{\,-1}
                \end{array}
      \right),
\\  \nonumber
\\ \nonumber
   \fl
 \left(\begin{array}{cc}
                \overline  A&i\overline B\\iB& -A
                 \end{array} \right)
 \left(\begin{array}{c}
                z\\ \overline z
                 \end{array} \right)=
 \left(\begin{array}{cc}
                \overline  A&i\overline B\\iB& -A
                 \end{array} \right)
\left(\begin{array}{cc}
                 -iB&A\\
                 \overline A&i\overline{B}
\end{array}\right)G^{-2}\left(\begin{array}{c}
                h\\ \overline h
                 \end{array} \right)=
\left(\begin{array}{cc}
                 0&I\\
                 -I&0
\end{array}\right) \left(\begin{array}{c}
                h\\ \overline h
                 \end{array} \right).
\end{eqnarray}
}

Note that  condition
\begin{equation}
\label{commute}
\fl
[A\overline A,B]=0,
\end{equation}
does not imply that $A$ and $B$ commute, but if  $A\overline A$ has a
simple spectrum, then $BA=A\overline B$. Indeed, since $[A\overline
A,B]=0$, the matrices $A\overline A$ and $B$ must have a joint system
of spectral projectors $\{ \widehat\pi_k\}$ such that
\begin{equation*}
\fl
A\overline A=\sum_k d^2_k \widehat\pi_k,\quad
B=\sum_k \lambda_k \widehat\pi_k,\quad \sum_k\widehat\pi_k=I,\quad \widehat\pi_k=\widehat\pi_k^*,\quad\widehat\pi_k\widehat\pi_j=I\delta_{kj},
\end{equation*}
where $d^2_k$ and $\lambda_k$ are the eigenvalues of $A\overline A$
and $B$ respectively, and $\widehat\pi_k$ are their common spectral
projectors. If all $\{d^2_k\}_1^n$ differ each other, then there exists
the polynomial
\begin{equation*}
\fl
f(d)=\sum_k\lambda_k\prod_{d_m\neq d_k}
\frac{d-d^2_m}{d^2_k-d^2_m}=\sum_kf_k d^k,
\quad f_k=f_k(\lambda,d)\in
{\mathbb C},\quad d\in{\mathbb R_+}
\end{equation*}
such that $f(d^2_k)=\lambda_k$, and $f(A\overline A)=B$ follows from
 $f(A\overline A)\widehat\pi_k=f(d^2_k)\widehat\pi_k=\lambda_k \widehat\pi_k$.  Therefore, the ``commutation relation''
\begin{equation}
\label{BA}
\fl
\overline A B=\overline A \sum_k f_k \,(A\overline A)^k=
\sum_k f_k \,(\overline A A)^k\,\overline A=\overline B \,\overline A
\end{equation}
is a consequence of \eref{commute} for matrices $A\overline A$ with
simple spectrum.

If the spectrum of  $A\overline A$  is multiple (for example,
$A=A\overline A=I$) and $B\neq\overline B$, then \eref{BA} clearly
fails. On the other hand, \eref{BA} holds for the operators $A$ such that
the  multiplicity of the spectrum of  $A\overline A$ is greater then or
equal to the spectral  multiplicity of $B$, because in such case the
polynomial representation $B=f(A\overline A)$ remains well-defined and
implies the equality $BA=A\overline B$ (see \cite{HoJo85}, sect.~4.4 for
applications of this equality  in linear algebra).

The relationship between the singular value decomposition of the
Hermitian matrix $A\overline A=U^* |D|^2 U$ (with unitary $U$ and
arbitrary diagonal matrix $D$),  and the general representation of the
symmetric matrix $A$ follows from a modified version of the Takagi
representation formula (see \cite{Ha07}): $A=U^*D\overline U$. In
order to satisfy \eref{commute}, we suppose that $B=U^*\Lambda U$
with the same unitary $U$ and arbitrary real diagonal matrix $\Lambda$.
Then $A$ is symmetric, $B$ is a Hermitian matrix,  $A\overline A=U^*D^2U$ and $B=U^*\Lambda U$
commute, and $A\overline B=B A=U^*\Lambda D\overline U$.

\section{Numerical tests for integral and algebraic representations of $s_t$}
Studying algebraic properties of the main objects
related to symplectic matrices \eref{defG}, we
have tested numerically non-trivial relations for randomly generated matrices $A$, $B$, and vectors $h$.
\begin{itemize}
\item[1.]  The following representations for $\gamma_t$ hold true:
{\small
\begin{eqnarray*}
\nonumber
\fl
\hskip-12mm
\gamma_t =
\int_0^tds \left(\left(\begin{array}{cc}
                 0&I\\
                  -I&0
\end{array}\right)\left(\begin{array}{c}
                 h_s\\
                  \overline h_s
\end{array}\right),\e^{Gs}
\left(
\begin{array}{c}
                 h\\
                 \overline h
 \end{array}
\right)\right)=
%\\
%\nonumber
%\fl
\left(\left(\begin{array}{cc}
                 0&I\\
                  -I&0
\end{array}\right)\left(\begin{array}{c}
                 h\\
                  \overline h
\end{array}\right),\left(\frac{I-Gt-e^{-Gt}}{G^{2}}\right)\,
\left(
\begin{array}{c}
                 h\\
                 \overline h
 \end{array}
\right)\right)
\\
%\fl
%\label{gamma}
=\left(\left(\begin{array}{cc}
                 0&I\\
                  -I&0
\end{array}\right)\left(\begin{array}{c}
                 h\\
                  \overline h
\end{array}\right),\left(\frac{e^{Gt}-Gt-I}{G^{2}}\right)\,
\left(
\begin{array}{c}
                 h\\
                 \overline h
 \end{array}
\right)\right).
\end{eqnarray*}
}

\item[2.] For $F_t=h_{-t}+R_t{\overline h}_{-t}$, the following  representations of the vacuum expectation
$\langle 0|e^{it\widehat H}|0\rangle=e^{s_t}$ are equivalent:
\begin{eqnarray}
\fl
\nonumber
e^{s_t}=e^{-\int_0^tds\bigl(\frac{1}{2}(\overline f_s,A\overline f_s)+\frac{1}{2}\tr(\overline{\rho}_sA)+(\overline{f}_s,h)\bigr)}=
e^{\int_0^tds\bigl(\frac{1}{2}(F_s,\overline AF_s)-\frac{1}{2}\tr(R_s\overline A)+(\overline h, F_s)\bigr)}=
\\
\label{vacAv}
\fl
\frac{e^{i\pi{\rm Ind}_t}}{\sqrt{\det\Phi_t}}\,
e^{\frac{1}{2}(\gamma_t-it\tr B+(\overline{h}_{-t},\Phi_t^{-1}h_t))}=
\frac{e^{i\pi{\rm Ind}_t}}{\sqrt{\det\Phi_t}}\,
e^{\frac{1}{2}(\gamma_t-it\tr B-(h_t,(\overline{h}_t-\rho^*_t h_t))}.
\end{eqnarray}
\item[3.]
Let $\alpha=2||(\Phi_t-\Psi_t){\rm Re\,}(\Phi_t-\Psi_t)^{-1}h_t||^2-{\rm Re\,}(g_t,(\Phi_t-\Psi_t)^{-1}h_t)$.
Then the unit norm of squeezed state $|A,B,h\rangle=e^{it\widehat H}|0\rangle$ can be equivalently represented in terms of various objects:
\begin{eqnarray}
\fl
\nonumber
1= || |A,B,h\rangle||^2=
\frac1{\sqrt{\det(I-\overline R_t R_t)}}\,
e^{\alpha_t-{\rm Re\,}\int_0^tds\bigl((\overline f_s,A\overline f_s)+\tr(\overline{\rho}_sA)+(\overline{f}_s,h)\bigr)}=
\\
\label{testNorm}
\fl
\frac1{\sqrt{\det(I-\overline R_t R_t)}}\,e^{2{\rm Re\,}(s_t)+\alpha_t}=
e^{{\rm Re\,}(\alpha_t+\gamma_t +(\overline{h}_{-t},\Phi^{-1}_t h_t))}=
e^{{\rm Re\,}(\alpha_t+\gamma_t -(\overline{h}_{t}+\overline{\rho}_th_t,h_t))}.
\end{eqnarray}
\end{itemize}
Note that the normalization conditions \eref{testNorm} are independent on the index function.

The graphs in fig.~1 and fig.~2 were created for randomly chosen $A$, $B$, and $h$:
{\small
\begin{eqnarray*}
\fl
A=\left(
\begin{array}{ccc}
 1.694 +0.3276 i & 0.317 +0.54 i & 0.509 +0.331 i \\
 0.317 +0.54 i & 0.0031 +1.9513 i & 0.6619 +0.0605 i \\
 0.509 +0.331 i & 0.6619 +0.0605 i & 0.5526 +0.5576 i
\end{array}
\right)
,\quad
h=\left(
\begin{array}{c}
 -0.6898+0.8259 i \\
 -0.3758+0.0629 i \\
 -0.4417-0.5016 i
\end{array}
\right)
\\
B=\left(
\begin{array}{ccc}
 1.3802 & 1.8946 +0.5657 i & 1.1696 +1.1702 i \\
 1.8946 -0.5657 i & 1.2728 & 1.7892 +1.3761 i \\
 1.1696 -1.1702 i & 1.7892 -1.3761 i & 0.5547
\end{array}
\right).
\end{eqnarray*}
}

The coincidence of $s_t$ in expressions  \eref{eqS} and \eref{ScalarFactor} was
tested for random matrices $A$, $B$, and the vector $h$ generated
numerically by using Wolfram Mathematica. For 3,\,4, and 5-modes
systems, the representation \eref{ScalarFactor} was implemented  650,
8000, and  141000  times faster; the values of $s_t$ calculated by \eref{eqS}
 and \eref{ScalarFactor}  coincide with accuracy $10^{-11}$.
The  functions  \eref{eqS} and \eref{ScalarFactor} are deposited at \cite{statphys} as Mathematica 7 modules.

\section*{Acknowledgments}
Authors thank Prof.  A.~S.~Chirkin for a useful discussion of the questions related to this paper.


\section*{References}

\begin{thebibliography}{99}

\bibitem{Be66} Berezin F A, The Method of Second Quantization, New
    York, 1966.

 \bibitem{Ki63} Kirznic D A, Field Theoretical Methods in Many-Body Theory, Gosatomizdat, Moscow, 1963.

\bibitem{Bo82} Bogoliubov N N, Shirkov D V, Quantum Fields, 1982,
    Mass., Benjamin-Cummings Pub. Co.

\bibitem{Ma65} Maslov V,  Theory of perturbations and asymptotic methods; appendix by V.~ I.~Arnold. French translation of Russian original (1965), Gauthier-Villars (1972).

\bibitem{Fa03}   Fan H Y,  2003 {\it J. Opt. B: Quantum Semiclass. Opt.},
    {\bf 5:4}  R147--R163.

\bibitem{GaZo04} Gardiner C W,  Zoller P,    Quantum Noise,
    Springer, Berlin, 2004.

\bibitem{He00} Haus H A, Electromagnetic noise and optical quantum measurements,  Springer, Berlin, 2000.

\bibitem{Do02} Dodonov V V, 2002 {\it  J. Opt. B: Quantum
    Semiclass. Opt.} {\bf 4} (1).

\bibitem{SuAg09} Sumei H,  Agarwal  G S, Squeezing of a
    Nanomechanical Oscillator,  {\it arXiv:0905.4234} [quant-ph].

\bibitem{ScCiWo06}  Schuch N, Cirac J I, and Wolf M M, 2006 {\it
    Commun. Math. Phys.} {\bf 267}, 65.

\bibitem{ChRaTl11} Chebotarev A M, Tlyachev T V, Radionov A A 2011
    {\it     Mathematical Notes}  {\bf 89} 577.

\bibitem{ChRaTl12} Chebotarev A M, Tlyachev T V, Radionov A A 2012
    {\it     Mathematical Notes} {\bf 92} 109.

\bibitem{Go13}    Maurice A. de Gosson, The symplectic egg in classical and quantum mechanics,  2013
   {\it  Am. J. Phys.} \textbf{81}, {\bf 328} 328-337.

\bibitem{Fe51} Feynman R P, 1951 {\it Phys. Rev.} \textbf{84}, 1, 108.

\bibitem{KaMa91}  Karasev M V and  Maslov V P, {\it Nonlinear Poisson
    Brackets: Geometry and Quantization}  [in
    Russian] Nauka, Moscow,  1991.

\bibitem{HoJo85} Horn R A, Johnson Ch R, Matrix Analysis, Cambridge
    Univ. Press  1985.

\bibitem{Ha07}  Hahn T, Routines for the diagonalization of complex
    matrices, arXiv:physics.comp-ph/0607103v2.

\bibitem{TlChCh13}  Tlyachev T V, Chebotarev A M, Chirkin A S, 2013,  {\it
    Physica Scripta T} {\bf 153} 014060.

\bibitem{statphys}  http://statphys.nm.ru/biblioteka/Demo/FactorS.nb

\bibitem{root} http://reference.wolfram.com/mathematica/tutorial/FunctionsThatDoNotHaveUniqueValues.html

\end{thebibliography}
\end{document}